\documentclass[a4paper,twocolumn,showpacs,superscriptaddress,floatfix]{revtex4-1}

\usepackage{latexsym}
\usepackage{amssymb}
\usepackage{amsfonts}
\usepackage{amsmath}
\usepackage{bm}
\usepackage{graphicx}   
\usepackage[squaren]{SIunits} 
\usepackage{color}
\usepackage{subfigure}
\usepackage{times}
\usepackage{units}
\usepackage{hyperref}
\usepackage{multirow}

\graphicspath{ {./plots/} }

\newcommand{\msec}{\usk\milli\second}

\newcommand{\Fref}[1]{Fig.~\ref{#1}}

\begin{document}

\title[short title]{Effects of Chiral Effective Field Theory Equation of State on Binary Neutron Star Mergers}

\author{Andrea Endrizzi}

\address{Physics Department, University of Trento, 
via Sommarive 14, I-38123 Trento, Italy}

\address{INFN-TIFPA, Trento Institute for Fundamental Physics
  and Applications, via Sommarive 14, I-38123 Trento, Italy}
  
\author{Domenico Logoteta}

\address{Dipartimento di Fisica ``E. Fermi'', Universit\`a di Pisa, Largo B. Pontecorvo 3, I-56127 Pisa, Italy}

\address{INFN, Sezione di Pisa, Largo B. Pontecorvo 3, I-56127 Pisa, Italy}

\author{Bruno Giacomazzo}

\address{Physics Department, University of Trento, 
via Sommarive 14, I-38123 Trento, Italy}

\address{INFN-TIFPA, Trento Institute for Fundamental Physics
  and Applications, via Sommarive 14, I-38123 Trento, Italy}

\author{Ignazio Bombaci}

\address{Dipartimento di Fisica ``E. Fermi'', Universit\`a di Pisa, Largo B. Pontecorvo 3, I-56127 Pisa, Italy}

\address{INFN, Sezione di Pisa, Largo B. Pontecorvo 3, I-56127 Pisa, Italy}

\author{Wolfgang Kastaun}

\address{Max Planck Institute for Gravitational Physics (Albert
  Einstein Institute), Callinstrasse 38, 30167 Hannover, Germany}

\address{Leibniz Universit\"at Hannover, Institute for Gravitational
  Physics, Callinstrasse 38, 30167 Hannover, Germany}

\author{Riccardo Ciolfi}

\address{INAF, Osservatorio Astronomico di Padova, Vicolo
  dell'Osservatorio 5, I-35122 Padova, Italy}

\address{INFN-TIFPA, Trento Institute for Fundamental Physics
  and Applications, via Sommarive 14, I-38123 Trento, Italy}

\date{\today}

\begin{abstract}
We present fully general relativistic simulations of binary neutron 
star mergers, employing a new zero-temperature chiral effective field
theory equation of state (EOS), the BL EOS. We offer a comparison
with respect to the older GM3 EOS, which is based on standard relativistic 
mean field theory, and separately determine the impact of the mass. We
provide a detailed analysis of the dynamics, with focus on the 
post-merger phase. For all models, we extract the gravitational wave
strain and the post-merger frequency spectrum. Further, we determine
the amount, velocity, and polar distribution of ejected matter, and 
provide estimates for the resulting kilonova signals. We also study the 
evolution of the disk while it is interacting with the hypermassive remnant, 
and discuss the merits of different disk mass definitions applicable before 
collapse, with regard to the mass remaining after black hole formation.
Finally, we investigate the radial mass distribution and rotation profile
of the remnants, which validate previous results and also corroborate a 
recently proposed stability criterion.
\end{abstract}


\maketitle


\section{Introduction}
\label{sec:intro}
The first detections of gravitational waves (GWs) by the advanced LIGO
network have officially started the era of GW
astronomy~\cite{LIGO:BBHGW:2016}. On August $1^{st}$ 2017 advanced
Virgo also joined the search for GW sources and on August $17^{th}$
2017 the first GW signal from a binary neutron star (BNS) coalescence was
detected~\cite{GW170817-detection}. The GW detection was followed by
a large number of electromagnetic counterparts observed by
ground- and space based telescopes \cite{GW170817-EM, Coulter:2017}. Those confirm
that (at least some) short gamma-ray bursts (SGRBs) are indeed
associated with binary neutron star mergers~\cite{GW170817-GRB} and
provided evidence that such systems are also the source of the
heaviest elements in the universe~\cite{GW170817-EM,
Metzger2017}. Unfortunately the current sensitivity of the Virgo and
LIGO detectors was not sufficient to detect a post-merger GW
signal~\cite{GW170817-postmerger} and therefore it is not clear if a
black hole (BH) or a neutron star (NS) were the result of the
merger. There are however ongoing analysis on the subject combining the GW emission with its EM counterparts. For example kilonova models and observations (e.g., see~\cite{Metzger:2014, Tanvir:2017,Margalit:2017}) seem to exclude a long-lived NS remnant and support the formation of a short-lived hypermassive NS after merger.
Moreover, if an NS, either hypermassive or supramassive, was
formed after the merger, an observation of its GW emission could have 
been used to put strong constraints on the equation of state (EOS) of NS
matter~\cite{Bauswein:2012:11101}. Nevertheless, preliminary constraints for the EOS were derived from tidal effects during the inspiral~\cite{GW170817-detection}. Note however that these 
constraints strongly rely on the accuracy of the waveform models. 
Further 
detectors, LIGO India and KAGRA, will be added to the GW detector network in the next years. 
Third generation detectors with higher sensitivities, 
such as the Einstein Telescope, are planned for the future. Such detectors may
increase the probability to detect post-merger GW emission as 
well as explore regimes which cannot be studied during the inspiral (e.g., study EOS at larger densities and temperatures~\cite{Radice2017}). 

BNS systems can be classified based on their total baryonic mass and
EOS. BNS systems with a total mass higher than the maximum mass that can be 
supported by uniform rotation (see~\cite{Stergioulas2003LRR}) 
could promptly form a BH after merger (if the mass 
is sufficiently high~\cite{Hotokezaka:2011:124008}) or form an hypermassive 
neutron star (HMNS) which can survive for up to ${\sim} 1$ s before collapsing to 
a BH.
BNS systems with a total mass below the maximum mass for uniformly rotating 
NSs, but above the maximum mass for non-rotating NSs, will instead
produce a supramassive neutron star (SMNS). A SMNS collapses
once gravitational waves and magnetic fields have decreased the angular momentum 
below a critical limit (which depends on mass and EOS).
A typical SMNS can therefore survive for hours or longer.
If the total mass of the BNS system is even below the maximum mass of a non-rotating NS,
the merger remnant is a stable NS~\cite{Giacomazzo2013ApJ...771L..26G}.

Two observations of ${\sim} 2\,\mathrm{M_\odot}$ NSs seem to indicate
that a significant fraction of BNS mergers could lead to the formation
of HMNS or SMNS after merger, depending on the NS EOS, since the most
common NS masses measured in galactic BNS systems systems are 
${\sim}1.35\, \mathrm{M_\odot}$~\cite{Piro2017}.

In this work we performed general relativistic hydrodynamic
simulations of equal mass BNS systems with different EOSs. For each, we chose initial values for the
NS masses which slightly exceed the maximum mass for a
uniformly rotating NS. The motivation is to study the HMNS case with different EOSs.

In the present paper we employed a new EOS for 
nuclear matter in $\beta$-equilibrium at zero temperature 
(BL EOS in the following) described in~\cite{BL2018}. 
It uses the Brueckner--Bethe--Goldstone (BBG) \cite{bbg1,BKL94,bbg2} many-body
theory with realistic two-body and three-body nuclear interactions
derived in the framework of chiral effective field theory (ChEFT).
This new EOS is used here for the first time to carry out general
relativistic hydrodynamic simulations of merging BNSs. 
For comparison, we also used a standard relativistic mean field EOS 
model for nuclear matter in $\beta$-equilibrium, specifically the GM3
parametrization of the Glendenning--Moszkowski \cite{gm91,glen00} EOS.
Note that in comparison to the BL EOS, the GM3 EOS leads to NS with
lower compactness and higher tidal deformability. Although the latter
approaches preliminary constraints obtained from GW170817~\cite{GW170817-detection}, 
we deemed this choice useful for studying EOS-related differences.

The paper is organized as follows. In Section \ref{sec:eos} we give a
more in-depth description of the new ChEFT EOS and explore the main
differences to the standard relativistic mean field one. In
Section \ref{sec:setup} we describe the initial data and the numerical
setup used for the simulations. In Section \ref{sec:dynamics} we
analyze the general dynamics of the BNS systems during inspiral and
post-merger, and the rotation profile of the remnants. In
Section~\ref{sec:kilonova} we discuss the EOS impact on the mass ejection 
and the consequences for the corresponding kilonova signal. In
Section~\ref{sec:gw} we present the GW signals and discuss the main
differences introduced by the new EOS. In Section \ref{sec:conclusion}
we present our main conclusions.

Unless specified otherwise we use a system of units in which
$G=c=M_{\odot}=1$.


\begin{figure}[t]
\includegraphics[width=0.5\textwidth]{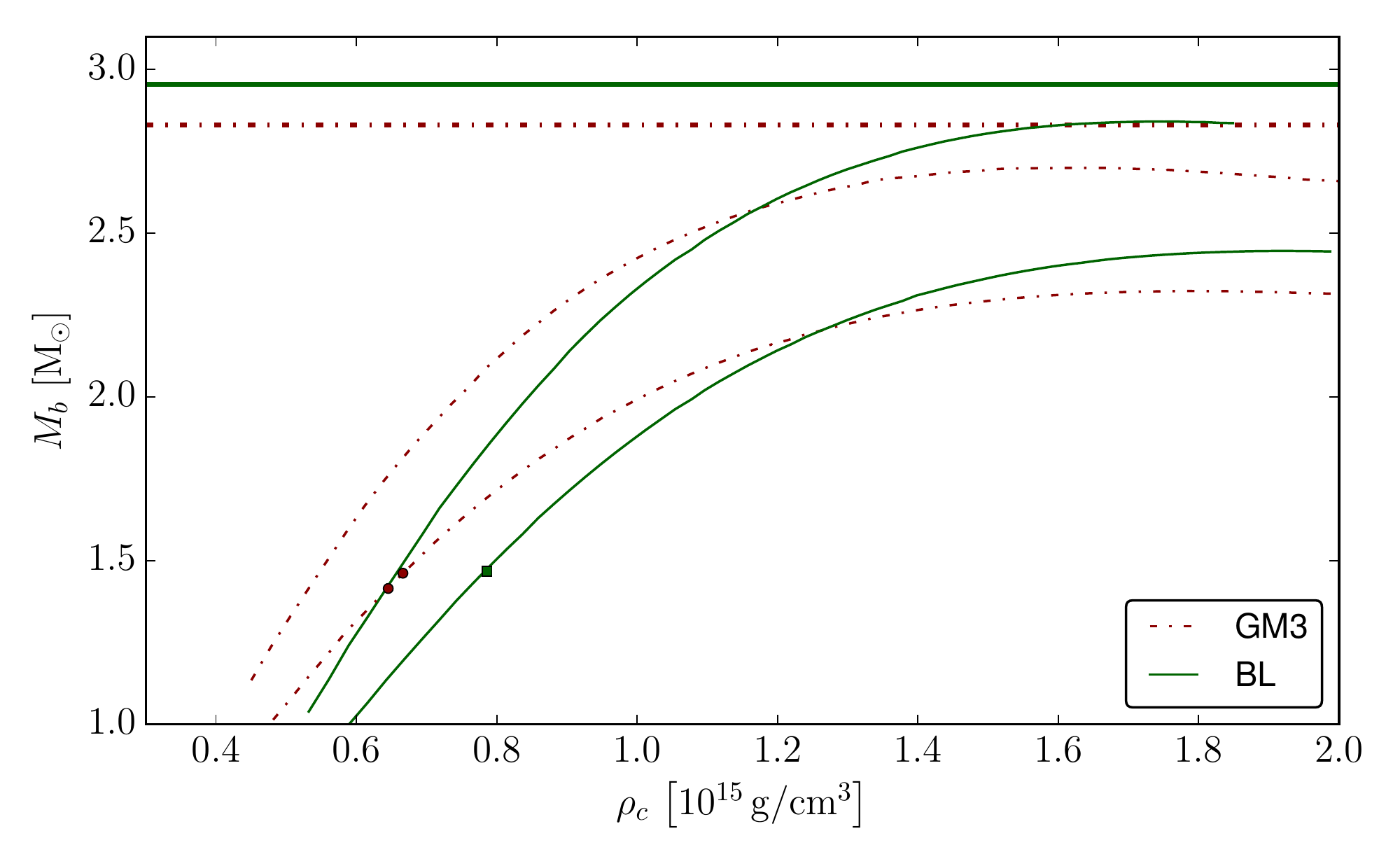}
\centering
\caption{Baryonic mass as a function of central rest mass density of non-rotating (lower curves) and maximally uniformly rotating (upper curves) NSs for the two EOSs used in this paper (BL and GM3). The horizontal lines represent the masses of the BNS systems, while the markers on the curves represent the mass values for the single NSs in the binary.
}
\label{fig:init_data}
\end{figure}

\begin{figure}[h]
\includegraphics[width=0.5\textwidth]{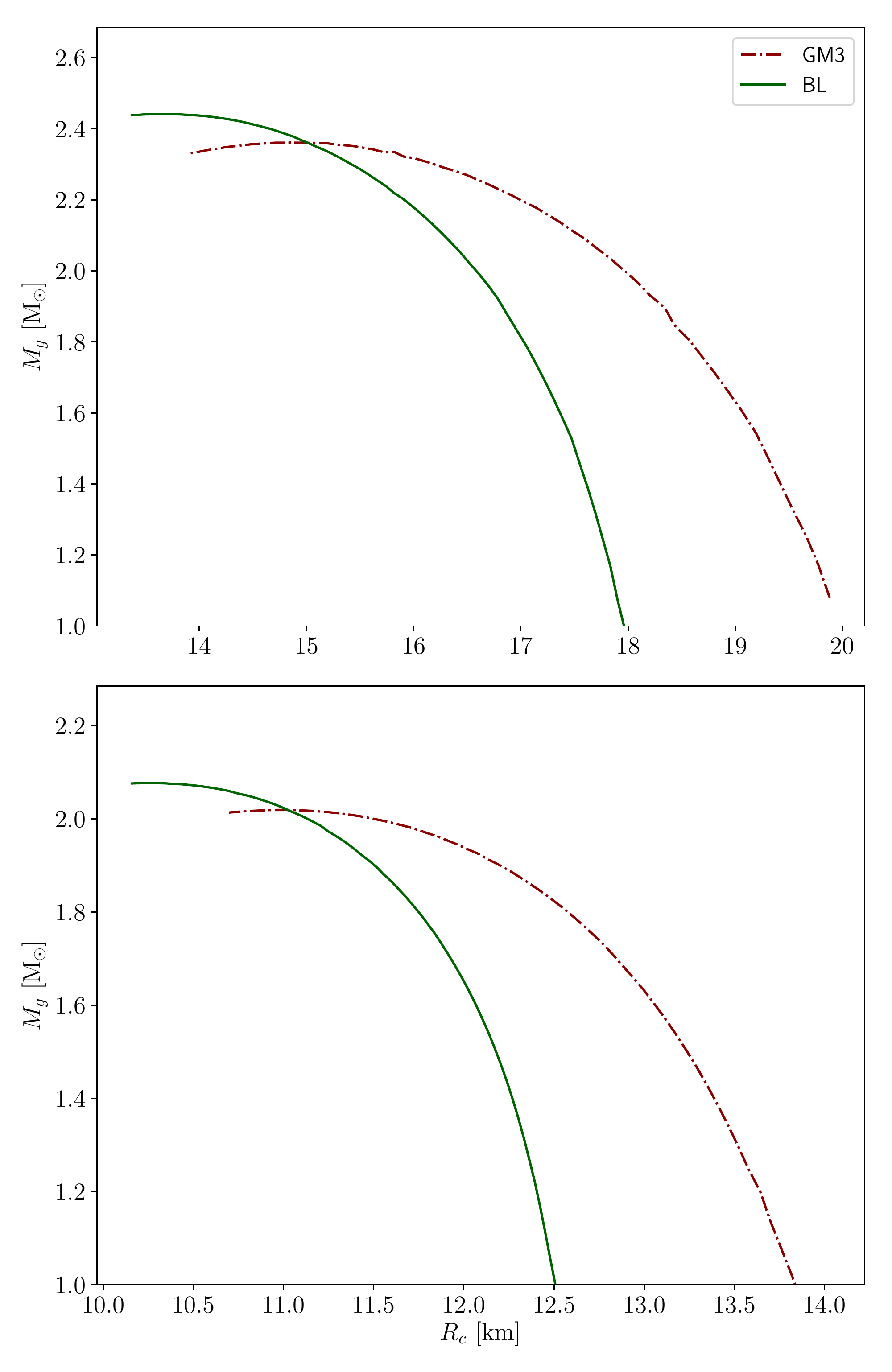}
\centering
\caption{Gravitational mass as a function of circumferential radius of NSs depending on the chosen EOS. 
Bottom panel: relation for non-rotating NSs. Top panel: relation for NSs rotating uniformly at mass shedding limit.
}
\label{fig:eos_mvr}
\end{figure}

\section{The BL Equation of State}
\label{sec:eos}
In the present work, we model the core of the two merging neutron stars as a uniform charge-neutral fluid of neutrons ($n$), protons ($p$), electrons ($e^-$) and muons ($\mu^-$) in equilibrium with respect to the weak interaction ($\beta$-stable nuclear matter). We neither consider the possible existence of ``exotic'' constituents in the stellar core, such as hyperons \cite{hyp1,hyp2}, nor the possibility of a deconfined quark phase \citep{glen00,bl2013,bombaci2016}.   

The BL EOS has been very recently derived in~\cite{BL2018}, making use of the the Brueckner--Bethe--Goldstone (BBG) quantum many-body theory in the Brueckner--Hartree--Fock (BHF) approximation (see e.g. \cite{LBK1,LBK2} and references quoted therein). The main innovative aspect of this EOS is the use of two-body and three-body nuclear interactions derived within the framework of the so called chiral effective field theory (ChEFT). This is a different approach with respect to other well known microscopic EOS models for $\beta$-stable nuclear matter, as e.g. the APR~\cite{apr98} EOS or the BBB~\cite{bbb97} EOS. In fact, ChEFT has opened a new route for the description of nuclear interactions \cite{weinberg1,weinberg2,weinberg3,weinberg4,epel09,machl11,holt13} consistent with quantum chromodynamics (QCD), the fundamental theory of the strong interaction. The most notable advantage of this method is that two-body, three-body, and even many-body nuclear interactions can be calculated perturbatively, i.e., order by order, according to a well-defined scheme.
The latter is based on a low-energy effective QCD Lagrangian which preserves the symmetries of QCD and particularly the approximate chiral symmetry. Within this chiral perturbation theory (ChPT) the details of the QCD dynamics are contained in parameters which are fixed by low-energy experimental data. This systematic technique is especially advantageous in the case of nuclear systems where the relevance of the three-nucleon force (TNF) is a well established feature. In fact, it is well known that TNFs are essential to reproduce the experimental binding energy of few-nucleon (A = 3, 4) systems \cite{kalantar12,hammer13,binder16} and the empirical saturation point ($n_{0} = 0.16 \pm  0.01~{\rm fm}^{-3}$, $E/A|_{n_0} = -16.0 \pm 1.0~{\rm MeV}$) of symmetric nuclear matter (SNM) \cite{FP81,apr98,bbb97,zuo14}. In addition, TNFs are crucial in the case of dense $\beta$-stable nuclear matter to obtain a sufficiently stiff EOS \cite{bbb97,apr98,li-schu_08} compatible with the measured masses, $M = 1.97 \pm 0.04 \, M_\odot$ \cite{demo10} and $M = 2.01 \pm 0.04 \, M_\odot$ \cite{anto13} of the neutron stars in PSR~J1614-2230 and PSR~J0348+0432 respectively. 

\begin{table*}[ht]
\caption{Properties of nuclear matter for the EOS models used in this work:   
saturation density $n_0$ and corresponding energy per nucleon $E/A$
for symmetric nuclear matter, as well as symmetry energy $E_{sym}^0$ and its slope parameter 
$L$ at the calculated saturation density. 
Further, $M_g^\mathrm{TOV}$ is the maximum gravitational mass of a non-rotating NS.
$M_g^\mathrm{SMNS}$ and $M_b^\mathrm{SMNS}$ are the maximum gravitational and baryonic 
masses, respectively, for a uniformly rotating NS.}
\begin{ruledtabular}
\begin{tabular}{cccccccc}
Model & $n_0$(fm$^{-3}$) & $E/A$ (MeV) & $E_{sym}^0$ (MeV)  & $L$ (MeV)  & $M_g^\mathrm{TOV}$ [$M_\odot$] & $M_g^\mathrm{SMNS}$ [$M_\odot$] & $M_b^\mathrm{SMNS}$ [$M_\odot$]\\
\hline
  BL & 0.171  & -15.23   & 35.39   &  76.0  & 2.08 & 2.44 & 2.84\\
 GM3 & 0.153  & -16.32   & 32.40   &  89.7  & 2.01 & 2.36 & 2.70\\
\end{tabular}
\end{ruledtabular}

\label{tab:tab_sat}
\end{table*}

In addition to the BL EOS, for the purpose of comparison, we also used a standard relativistic mean field EOS (GM3~\cite{gm91,glen00}). To model the NS crust we used the EOS reported in Ref.~\cite{HP1994} for the outer crust and the EOS reported in Ref.~\cite{DH2001} for the inner crust. 

The calculated values of the saturation points of SNM for the BL and GM3 EOS models are reported in Tab.~\ref{tab:tab_sat}. Both models give saturation points in very good agreement with the empirical ones. In addition, the calculated gravitational maximum mass $M_g^{TOV}$ for non rotating NS (see. Tab.~\ref{tab:tab_sat}) for the two EOS models is almost the same (relative difference $\sim 3 \%$) indicating that the overall stiffness of the two EOS is very similar. Obviously this result does not imply that the two EOS models will give similar results for other neutron star properties (e.g. mass-radius relation, moment of inertia, mass shed frequency, tidal deformability, etc) or in binary neutron star mergers simulations. 

The nuclear symmetry energy $E_{sym}$ \cite{epja50,baldo16}, and particularly its density dependence, is another important physical quantity which influences the properties of asymmetric nuclear matter (i.e. matter with $n_n \neq n_p$, where $n_n$ and $n_p$ denote the neutron and proton number densities, respectively). In particular, the symmetry energy determines the proton fraction and the pressure of nuclear matter in $\beta$-equilibrium \citep{bl91,zuo14}. Consequently, it has an impact on many NS attributes such as the radius, the moment of inertia and the tidal deformability \citep{latt14,LattPrak2001}.  

The symmetry energy can be obtained \cite{bl91} taking the difference between the energy per nucleon $E/A$ of pure neutron matter and the one of SNM at a given total nucleon number density $n = n_n + n_p$. The symmetry properties of nuclear matter around the saturation density $n_0$ are summarized by the value of $E_{sym}^0 \equiv E_{sym}(n_0)$ and by the value of the so called symmetry energy slope parameter 
\begin{equation}
       L = 3 n_{0} \frac{\partial E_{sym}(n)}{\partial n}\Big|_{n_{0}} 
\label{slope}
\end{equation} 

It has been shown \cite{latt2013} that a strong correlation between the values of $E_{sym}^0$ and $L$ can be deduced in a nearly model-independent way from nuclear binding energies. In addition, it has been recently demonstrated \cite{tews2017} that the unitary gas limit \cite{zwier2015}, which can be used to describe low density neutron matter, puts stringent constraints on the possible values of the symmetry energy parameters, excluding a large region in the $E_{sym}^0$--$L$ plane (see Fig. 3 in Ref.~\cite{tews2017}).     
As pointed out by the authors of Ref.~\cite{tews2017}) several tabulated EOS currently used in astrophysical simulations of supernova explosions and BNS mergers violate the unitary gas bounds. Thus the unitary gas model can be used as a novel way to constrain dense matter EOS to be used in astrophysical applications.  
  
The values of $E_{sym}^0$ and $L$ calculated for the BL and GM3 EOS models are reported in Tab.~\ref{tab:tab_sat}. These values are compatible with the unitary gas bound given in Ref.~\cite{tews2017} for the case of the BL EOS whereas they are not compatible  with the unitary gas bound in the case of the GM3 EOS.

The differences between the two EOSs in terms of mass versus rest mass density relation are shown in Fig \ref{fig:init_data}. As one can see, the ChEFT EOS can support a higher maximum mass than the GM3 EOS. Both EOSs allow NSs with gravitational mass above  $2 M_\odot$, compatible with current observational lower limits on the maximum mass~\citep{demo10, anto13}.
In Fig.~\ref{fig:eos_mvr}, we show the gravitational mass versus radius for maximally uniformly rotating  and non-rotating
NSs described by the EOSs used in this paper. 
The BL EOS leads to more compact stars than the GM3 EOS: for a NS with $M_g = 1.35 M_\odot$, the radius is around 
$1\usk \mathrm{km}$ smaller.

The BL and GM3 EOS do not include thermal effects. 
For the evolution, we add a thermal component using a standard gamma-law prescription, 
described in Appendix~\ref{sec:Tfeat}. 
Such hybrid EOSs are often used in the field of BNS simulations, 
see~\cite{Kiuchi:2014:41502,Bauswein:2010:84043}.

Note that the EOS assume $\beta$-equilibrium, and do not provide 
composition dependency out of $\beta$-equilibrium.
However, the matter will not stay in $\beta$-equilibrium during the rapid evolution at merger.
While the density changes rapidly, the electron fraction changes on
timescales given by weak processes. The presence of trapped neutrinos
in high density regions, and particularly their influence on the
matter composition (particle fractions) and their contribution to the
pressure further complicates the picture~\cite{Bombaci1996,Prakash1997}.
By using the BL or GM3 EOS, we will make
the implicit approximation that $\beta$-equilibrium holds during merger.


\section{Initial data and numerical setup}
\label{sec:setup}
In this work we evolve an unmagnetized equal-mass BNS system 
(with gravitational masses $1.35 M_{\odot}$) employing the new BL EOS. 
We also study two BNS systems with the older GM3 EOS. 
One has the same gravitational masses as the BL model. 
The other is 3\% lighter, such that the ratio 
$M_\mathrm{b}^\mathrm{tot} / M_\mathrm{b}^\mathrm{SMNS}$
is closer to the value for the BL model. All models 
have a total mass $4$--$8\%$ larger than the maximum possible for uniformly 
rotating models. We therefore expect that the mergers result in short lived 
remnants, and not in prompt collapse (compare \cite{Bauswein:2013:131101}).
The three models allow to determine separately the influence of the EOS
and of the mass. Their parameters are summarized in Table~\ref{tab:init_param}.
 
\begin{table*}
  \caption{Initial data parameters. $M_\mathrm{b}^\mathrm{SMNS}$  is the maximum 
    baryonic mass for a uniformly rotating NS,    
    $M_\mathrm{b}^\mathrm{tot}$  is the total baryonic mass of the system, 
    $M_\mathrm{g}$ and $R_c$ are gravitational mass 
    and circumferential radius of each star at infinite separation, 
    $f_0$ is the initial orbital frequency, $d$ the initial proper
    separation, and $\Lambda/M_g^5$ is the dimensionless tidal deformability (see~\cite{Hinderer:2010:123016}). 
     }
  \begin{ruledtabular}\begin{tabular}{lllllll}
      Model & BL-1.35 & GM3-1.31 & GM3-1.35\\
      \hline
      $M_\mathrm{b}^\mathrm{tot}$ [$M_\odot$]  & $2.95$  & $2.83$ & $2.92$\\
      $M_\mathrm{b}^\mathrm{tot}/M_\mathrm{b}^\mathrm{SMNS}$ & $1.04$ & $1.05$ & $1.08$\\
      $M_\mathrm{g}$ [$M_\odot$]        & $1.35$  & $1.31$ & $1.35$\\
      $M_\mathrm{g}/R_c$                  & $0.162$  & $0.144$ & $0.148$\\
      $f_0$ [Hz]                 & $281$  & $278$ & $282$\\
      $d$ [km]                   & $58$  & $57$ & $58$\\
      $\Lambda / M_g^5$			& $492$ & $957$ & $793$\\
  \end{tabular}\end{ruledtabular}
   \label{tab:init_param}
\end{table*}

Initial data are computed with the publicly available \texttt{LORENE} code~\cite{Gourgoulhon:2001:64029, Taniguchi2002}, 
assuming an irrotational binary on a quasi-circular orbit. 
We note that the initial absence of radial velocity leads to some eccentricity during the evolution.
The initial proper separation of $57-58\ \mathrm{km}$ allows the completion of 5--6 orbits.

All the simulations discussed in this paper use the {\tt WhiskyThermal} code~\cite{Galeazzi:2013:64009,Alic:2013:64049}
for the evolution of the hydrodynamic equations in general relativity. 
The spacetime is evolved with the publicly available {\tt McLachlan} code, which implements the BSSNOK formulation of the field equations~\cite{Baumgarte:1998:24007,Shibata:1995:5428,Nakamura:1987:1}. 
Both parts are coupled together within the publicly available 
Einstein Toolkit~\cite{Loeffler:2012:115001}. The latter also includes the {\tt Carpet}
code~\cite{Schnetter:2004:1465},  which we use for adaptive mesh refinement.

The general relativistic hydrodynamic equations are evolved using high resolution shock-capturing schemes based on the flux-conservative ``Valencia'' formulation~\cite{Banyuls:1997:221}. We compute fluxes using the HLLE \cite{Harten:1983:35} approximate Riemann solver, which uses primitive variables reconstructed at cell interfaces with the piecewise parabolic method. Finally we employ an artificial atmosphere with a floor value of $\rho_\mathrm{atmo} = 6.2\times 10^4\ \mathrm{g\ cm}^{-3}$ as well as zero velocity and temperature.

For the mesh refinement, we use fixed-size moving boxes during inspiral. The two refinement levels with the highest resolutions consist of nested cubes following the two NSs. Four additional coarser levels consist of larger, non-moving cubes. At merger, we switch to fixed mesh refinement. The cube of highest resolution has a half-diameter of $26\ \mathrm{km}$, covering
the remnant and the innermost part of the accretion disk.
The finest grid resolution is $dx = 186\ \mathrm{m}$, which means the NSs are resolved with ${\sim} 53-60$ points per coordinate radius, depending on the model.
The outer boundary in our simulations is located at $1178\ \mathrm{km}$. In order to save computational resources, 
we evolve the system with reflection symmetry across the orbital plane.


\section{Merger and postmerger dynamics}
\label{sec:dynamics}
In the following, we describe the dynamics of inspiral and merger, and 
investigate the evolution of the remnant properties.
The quantitative results discussed below are summarized in 
Table~\ref{tab:outcome}.

\begin{table*}
\caption{Outcome of our BNS mergers. $M_\mathrm{BH}$ and 
  $J_\mathrm{BH}$ are black hole mass and angular momentum $5\msec$ 
  after formation (only for collapsing models). 
  $M_\mathrm{blk}$ and $R_\mathrm{blk}$ are bulk mass and bulk 
  radius (see text for definitions), while $\nu_\mathrm{c}$ and 
  $\nu_\mathrm{max}$ denote the remnants central and maximum rotation 
  rates, all computed $12\usk\milli\second$ after merger.
  $f_\mathrm{merge}$ is the gravitational wave instantaneous frequency 
  at the time of merger, $f_\mathrm{pm}$ is the frequency of the maximum 
  in the post-merger part of the gravitational wave power spectrum, and
  $f_{10}$ is the amplitude-weighted average instantaneous frequency 
  during the first $10\msec$ after merger (see \citep{Ciolfi2017}). 
  $M_\mathrm{disk}^\mathrm{alt}$ is the mass at densities below 
  $10^{-13}\,\mathrm{g}/\mathrm{cm}^3$, $M_\mathrm{disk}$ is the mass 
  outside a coordinate sphere of volumetric radius $r=30$ km, 
  $M_\mathrm{fb}$ is the bound mass outside coordinate radius 
  $r>60$ km, all measured $12\usk\milli\second$ after merger. The value 
  in brackets denotes the mass outside the apparent horizon at 
  $5\msec$ after BH formation. 
  Finally, $M_\mathrm{ej}$ and $v_\infty$ are our estimates for the 
  total ejected mass and the average expansion velocity at infinity.}
\begin{ruledtabular}
\begin{tabular}{lccc}
Model&
BL 1.35&
GM3 1.31&
GM3 1.35
\\\hline 
$M_\mathrm{BH}\,[M_\odot]$&
---&
---&
$2.54$
\\
$J_\mathrm{BH}/M^2_\mathrm{BH}$&
---&
---&
$0.65$
\\
$M_\mathrm{blk}\,[M_\odot]$&
$2.52$&
$2.40$&
$2.52$
\\
$M_\mathrm{blk}/R_\mathrm{blk}$&
$0.30$&
$0.26$&
$0.29$
\\
$\nu_\mathrm{c}\, [\mathrm{kHz}]$&
$0.79$&
$0.68$&
$0.72$
\\
$\nu_\mathrm{max}\, [\mathrm{kHz}]$&
$1.59$&
$1.26$&
$1.43$
\\
$f_\mathrm{merge}\, [\mathrm{kHz}]$&
$1.87$&
$1.72$&
$1.69$
\\
$f_\mathrm{pm}\, [\mathrm{kHz}]$&
$3.17$&
$2.74$&
$2.89$
\\
$f_{10}\, [\mathrm{kHz}]$&
$3.09$&
$2.66$&
$2.79$
\\
$M_\mathrm{disk}\,[M_\odot]$&
$0.086$&
$0.077$&
$0.057$ ($0.050$)
\\
$M_\mathrm{disk}^\mathrm{alt}\,[M_\odot]$&
$0.136$&
$0.135$&
$0.097$
\\
$M_\mathrm{fb}\,[M_\odot]$&
$0.062$&
$0.046$&
$0.040$
\\
$M_\mathrm{ej}\, [10^{-2} \, M_\odot]$&
$0.62$&
$0.14$&
$0.10$
\\
$v_\infty\, [c]$&
$0.17$&
$0.13$&
$0.10$
\end{tabular}
\end{ruledtabular}
\label{tab:outcome}
\end{table*}

\begin{figure}[t]
\includegraphics[width=0.5\textwidth]{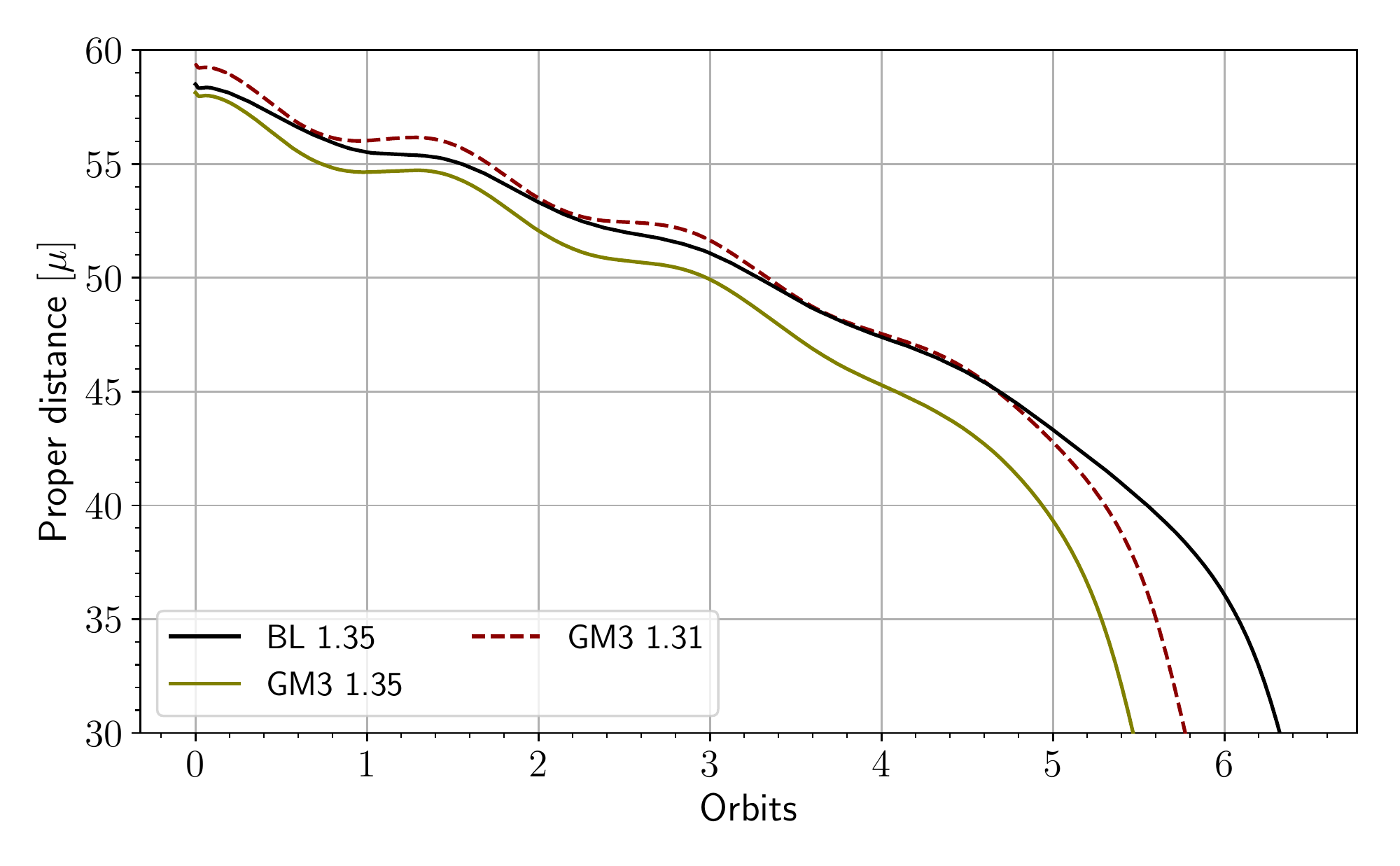}
\centering
\caption{Proper separation between the NS barycenters during the 
inspiral as function of orbital phase.
The separation is given in units of reduced mass $\mu =M_g^1 M_g^2 / (M_g^1 + M_g^2)$. 
The barycenters and orbital phases are computed using simulation coordinates.}
\label{fig:prop_sep}
\end{figure}

All three BNS models complete ${\sim} 5$--$6$ orbits before merger. 
Fig.~\ref{fig:prop_sep} shows the proper separation as function of the orbital phase. 
We find a residual eccentricity which is typical for initial data obtained with the
quasi-circular approximation (see Section \ref{sec:setup}) without corrections for 
the radial velocity of the inspiral.
Therefore, we do not try to make quantitative statements about tidal effects on the 
orbital dynamics. Qualitatively, we find a longer inspiral for model $BL-1.35$.
Eccentricity aside, this agrees with the theoretical expectation because 
the NSs are more compact than those of the GM3 models.

None of our models forms a BH directly at merger. Consequently, the threshold 
for prompt collapse (in the equal mass case) is above a total gravitational mass of 
$2.7\,M_\odot$ for both EOS. 
For model $GM3-1.35$, a BH is formed ${\sim} 14$ ms after merger (final mass
and angular momentum given in Table~\ref{tab:outcome}). 
For the other two models, $GM3-1.31$ and $BL-1.35$, the remnants survive at least 
$5\usk\milli\second$ longer, at which point our simulations end.
A strong influence of EOS and mass on the lifetime is not surprising,
since our systems are in the hypermassive mass range.

\begin{figure*}[t]
\includegraphics[width=0.95\textwidth]{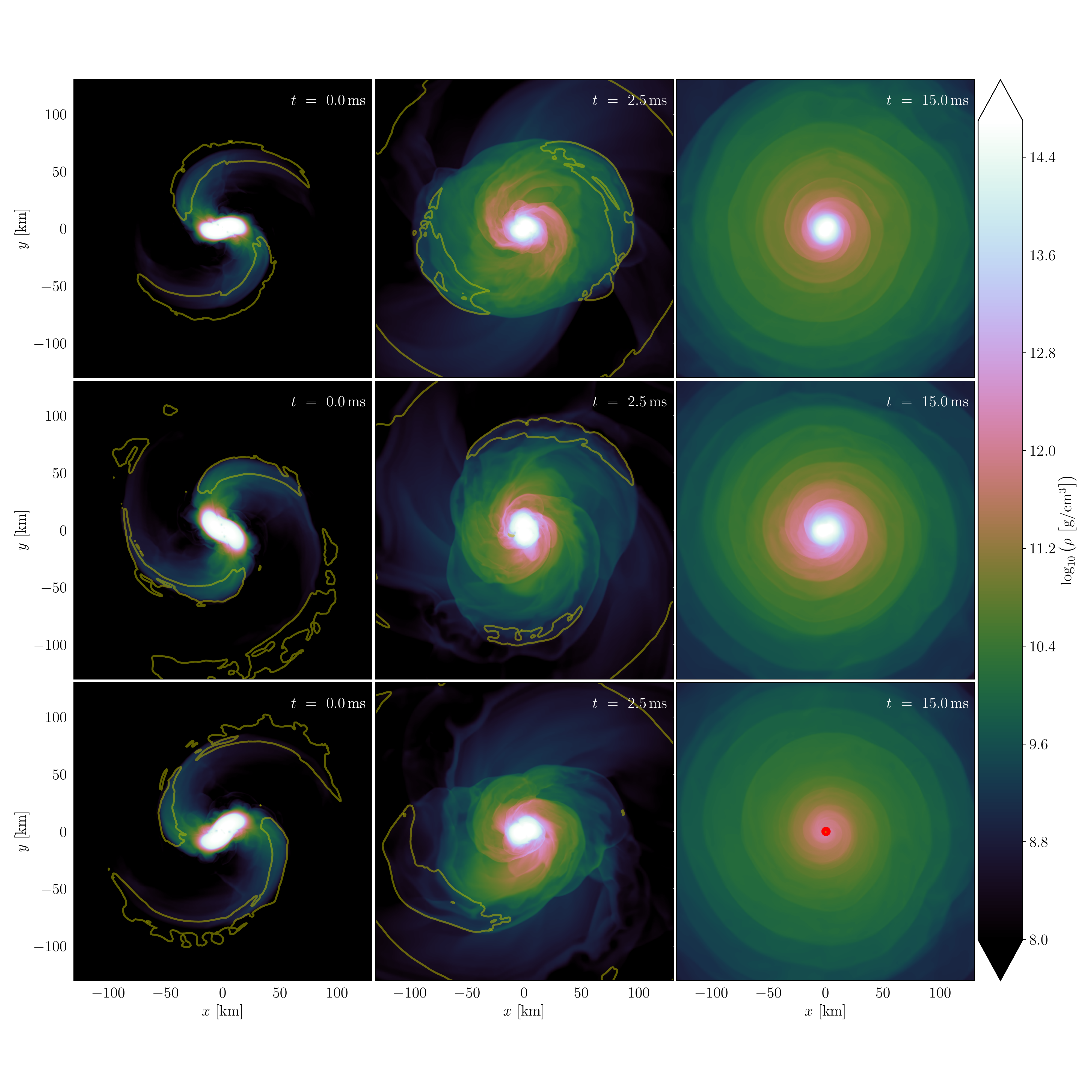}
\centering
\caption{Rest-mass density evolution in the equatorial plane for the three simulations $BL-1.35$ (top), $GM3-1.31$ (middle) and $GM3-1.35$ (bottom). The yellow contour lines highlight the unbound ejected matter. The snapshots are taken at three different phases of the evolution: on the left at the time of merger, in the middle $2.5$ ms after merger, and on the right $15$ ms after merger. The apparent horizon of the newly formed BH for the $GM3-1.35$ model is highlighted with a red circle.}
\label{fig:rho_xy}
\end{figure*}

\begin{figure*}[t]
\includegraphics[width=0.95\textwidth]{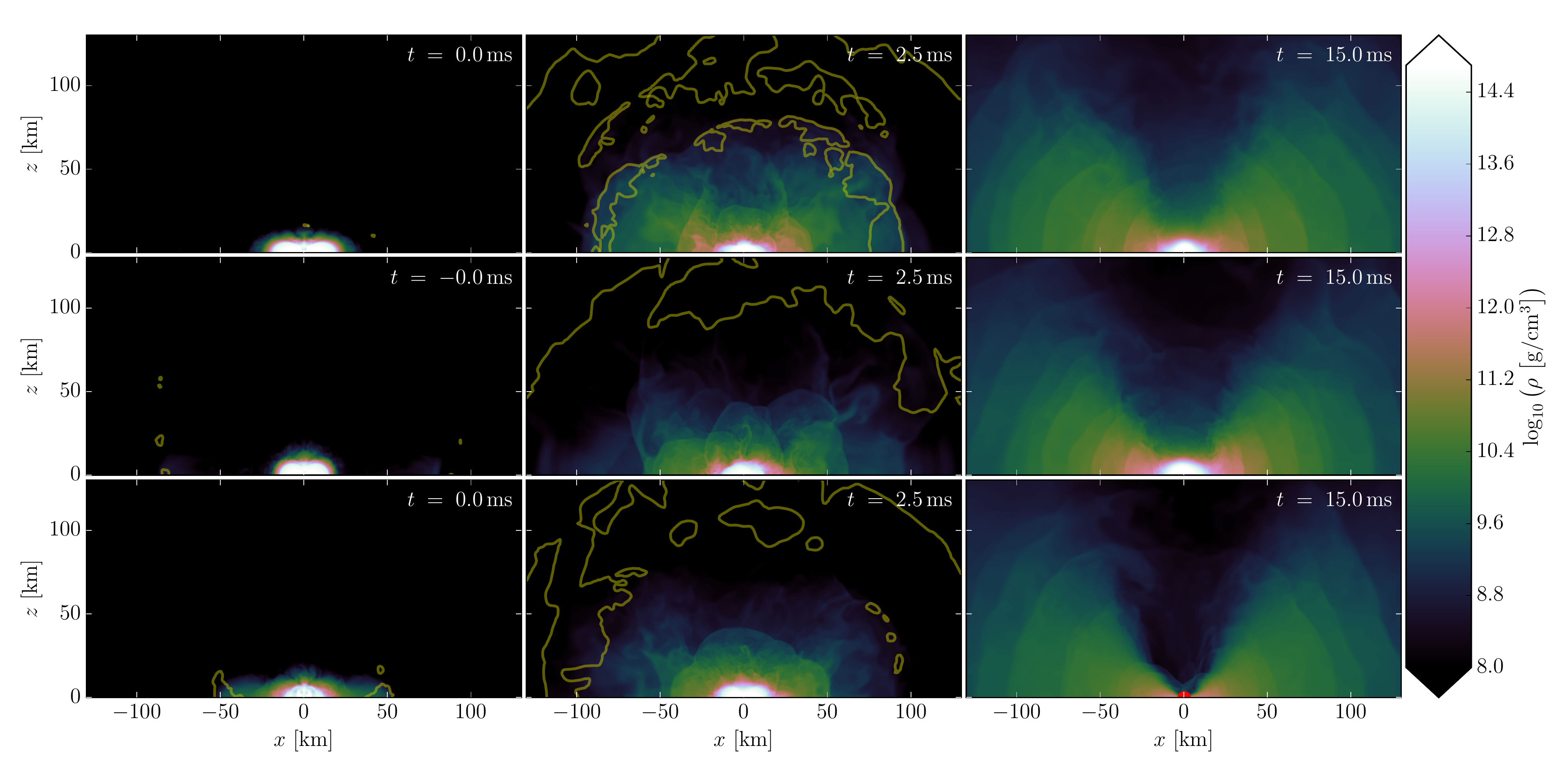}
\centering
\caption{Same as Fig.~\ref{fig:rho_xy}, but showing a cut in the meridional plane.} 
\label{fig:rho_xz}
\end{figure*}

Snapshots of the density distribution after merger are given in 
Figures~\ref{fig:rho_xy} and~\ref{fig:rho_xz} for all three models. 
One can observe the formation of a disk in all cases.
It is worth noting that the disk keeps growing until a BH forms.
The cause might be a complex fluid flow in the outer 
layers of the strongly deformed remnant, involving rotational vortices. 
Such a mechanism was investigated in detail for a different model in 
\cite{Kastaun:2016}. 
Also the temperature evolution (based on the simplified prescription for 
the thermal pressure) is qualitatively similar to the one in \cite{Kastaun:2016},
as shown in Appendix~\ref{sec:Tfeat}.

\begin{figure}[t]
\includegraphics[width=0.5\textwidth]{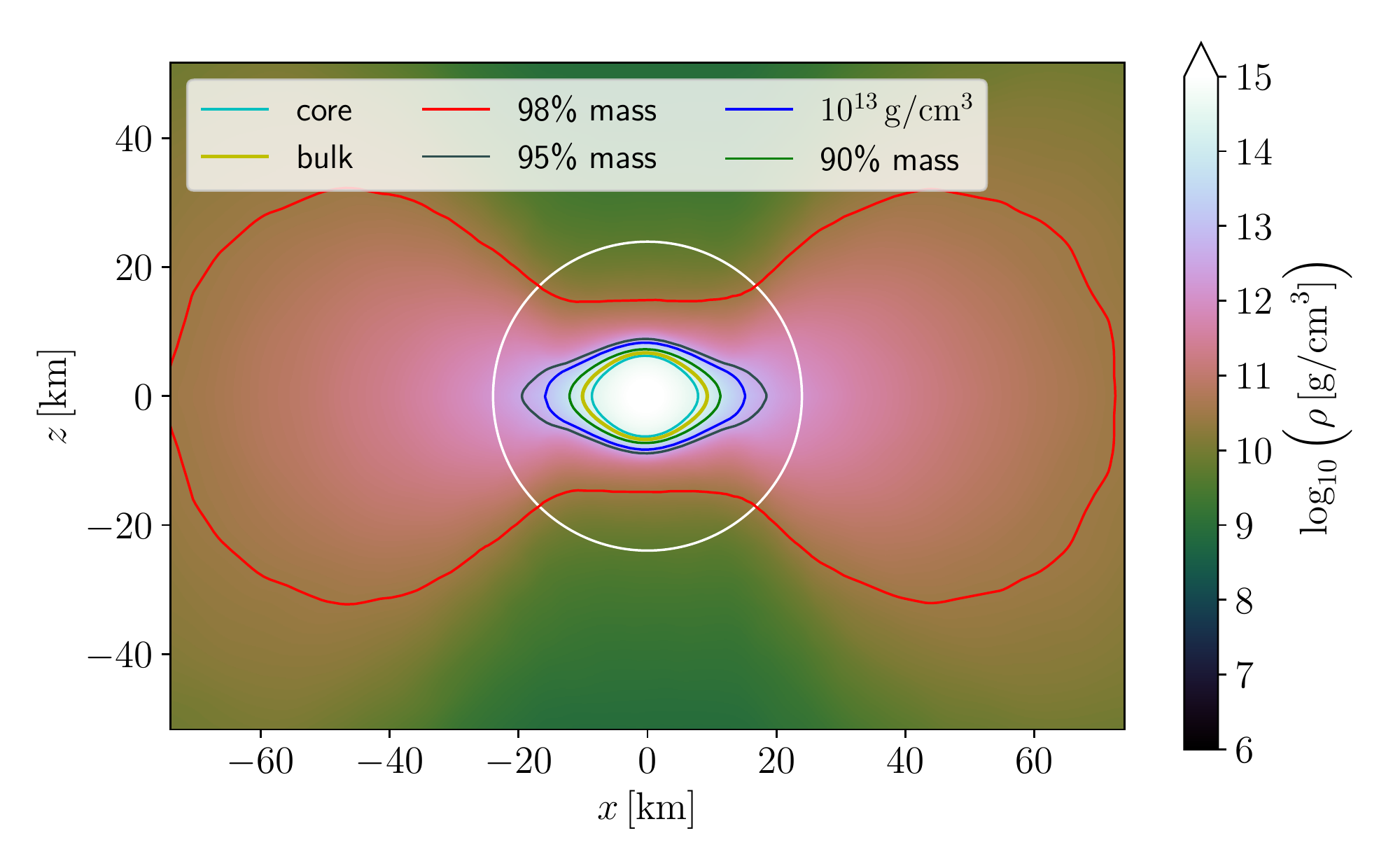}
\centering
\caption{Structure of the disk and embedded remnant for model $GM3-1.31$, at the
end of the simulation.
The mass density is shown as color plot. In order to remove contributions from 
oscillations, we average in time over the last $4\usk\milli\second$.
The white circle marks the cutoff (coordinate) radius used to define the disk mass $M_\mathrm{d}$.
The density cutoff $10^{13}\, \mathrm{g/cm}^3$ used for the alternative definition 
$M_\mathrm{d}^\mathrm{alt}$ is marked by the blue contour. The other contours mark
isodensity surfaces containing $90,95$, and $98\%$ of the total baryonic mass, and
those corresponding to remnant bulk and core (see text).}
\label{fig:gm3_disk}
\end{figure}

In order to quantify the properties of the disk, we first need
to discuss its structure, in particular the lack of a clear distinction
between a HMNS and the surrounding disk. Fig.~\ref{fig:gm3_disk} shows 
one example of the disk structure. 
Ideally, we would like to distinguish between remnant and disk mass in 
a meaningful way which is also easy to extract from the simulation data.
From the available data, we can extract the total mass outside a given cutoff 
radius or below a certain density threshold. However, such measures depend strongly
on the chosen cutoff values. This can be seen from the density isosurfaces
shown in Fig.~\ref{fig:gm3_disk} and the enclosed mass fractions (given in the legend). 

One meaningful definition of a disk (at a given time) is the region
which would not be swallowed if the central remnant would suddenly collapse 
to a BH. This will depend mainly on mass and spin of the BH. 
For model $GM3-1.35$, we picked a cutoff radius such that the mass
outside shortly before collapse roughly corresponds to the mass remaining outside 
of the apparent horizon shortly after collapse. 
Since we expect similar properties of the final BH for the other models, we use 
the same definition for the disk mass $M_\mathrm{d}$ also for these. To reduce 
gauge ambiguities, we specify the cutoff radius based on the proper volume enclosed
inside the coordinate sphere. The chosen cutoff volume is that of an Euclidian sphere 
with a radius of $30\usk\kilo\meter$, the corresponding coordinate radius is marked
in Fig.~\ref{fig:gm3_disk} with a white circle.

For comparison with \cite{Radice2018}, we also employ an alternative measure for the 
disk mass, $M_\mathrm{d}^\mathrm{alt}$, defined as the total mass below a 
density of $10^{13}\usk\gram\per\centi\meter\cubed$. As shown in 
Fig.~\ref{fig:gm3_disk}, this definition cuts the disk much closer 
to the remnant. In particular, it includes material around the rotation axis, which 
will be swallowed when the BH is formed. At least for our models, the resulting disk mass is not a good estimate for the disk mass remaining after collapse. Similarly, it might depend more strongly on the exact structure of the remnant. This is relevant for the interpretation of \cite{Radice2018}, where the measure is used as upper limit for mass ejection from the disk for the cases where no BH forms during the simulation.

\begin{figure}[t]
\includegraphics[width=0.5\textwidth]{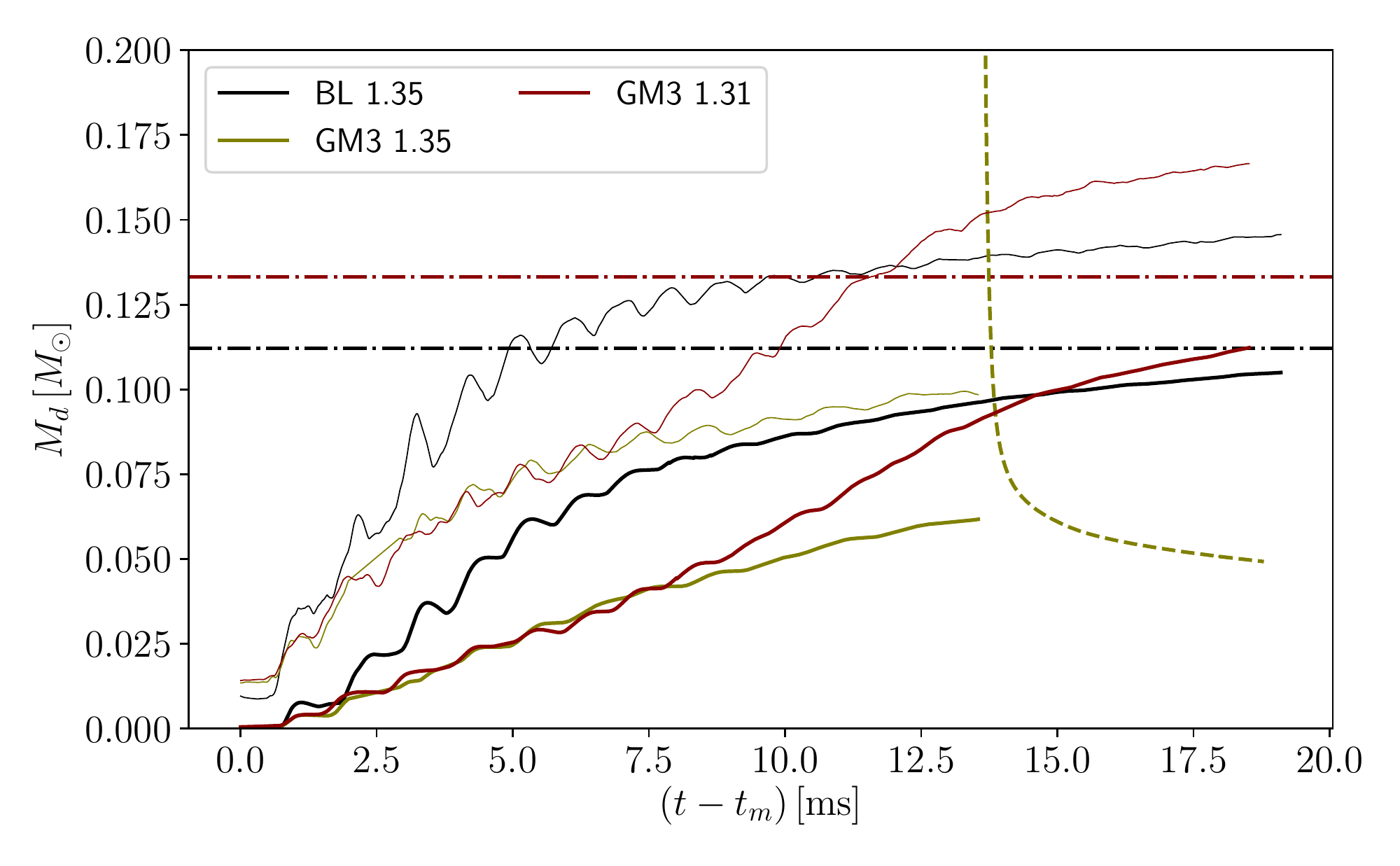}
\centering
\caption{Time evolution of disk mass for our simulations.
The solid thick lines show the disk mass measure $M_\mathrm{d}$, and
the thin solid lines the alternative measure $M_\mathrm{d}^\mathrm{alt}$.
After the BH formation for model $GM3-1.35$, we show the mass outside
the apparent horizon instead (dashed line). The rapid initial decrease 
corresponds to the rapid initial growth of the apparent horizon.
The horizontal lines mark a disk mass for which the remaining remnant 
mass is at the upper limit for a SMNS (the two GM3 models are 
indistinguishable due to almost identical total mass).
}
\label{fig:evol_mdisk}
\end{figure}

The time evolution of the disk masses is shown in Fig.~\ref{fig:evol_mdisk}.
We can see that the two disk definitions differ significantly, 
by a factor ${\approx}1.6$. Note that the error of the density based estimate
will increase when the true value decreases since the estimate includes
some material that will collapse together with the remnant.
As mentioned before, the disk is increasing gradually before BH formation.
Such a mass expulsion might delay the collapse of HMNSs and also contribute to
the seemingly unpredictable amplitude evolution of GW signals from HMNSs.

Note that the values we provide in 
Table~\ref{tab:outcome} refer to a specific time.
Although the growth rate slows down, the disk masses are not constant.
We stress that values for the disk mass extracted from numerical simulations
should be regarded only as ballpark figures because of the ambiguities arising 
from extraction time and disk mass definition.  

Comparing our results for disk and ejecta masses to Fig.~1 
in \cite{Radice2018}, we find that models $BL-1.35$ and $GM3-1.35$ are outliers 
to the proposed correlation. Although the disk masses
in \cite{Radice2018} for non-collapsing models were extracted somewhat later 
than ours, our disk mass $M_\mathrm{d}^\mathrm{alt}$ for $BL-1.35$ is larger 
and still increasing at the end of our simulation. For $GM3-1.35$, the mass 
remaining outside the BH is lower and still decreasing.
The deviation might be due to the model selection in \cite{Radice2018}, or due 
to the numerical errors of our simulation. Further high-resolution studies are 
needed to estimate the latter. Nevertheless, our results indicate that the ambiguities 
in disk mass definition and extraction time add a factor ${\approx}2$ to the disk mass error.

We also point out that the mass fraction which is ejected from the disk for a given model
is another important unknown not considered in \cite{Radice2018}. 
Studies modeling said fraction
find typical values in the range $10-50\%$ with different methods and
assumptions (see e.g.,~\cite{Metzger2008, Metzger:2009, Metzger:2014, 
Fernandez:2015:nsns, Just:2015, Siegel:2018}). If, for example, only $1/5$ of the 
disk mass can be ejected, none of the simulations would reach the required ejecta 
mass. The above references also indicate that
a reliable measure for the mass of the disk might be required, but not
sufficient, because disk structure and neutrino irradiation, as well as
magnetic fields, also play an important role for the ejected mass.
In order to constrain the EOS this way, a reliable modeling of winds ejected 
from the disk is required in addition to robust kilonova lightcurves models.

For both measures, the disk mass is initially larger for model $BL-1.35$ than 
for the GM3 models. Interestingly, the growth rate of the disk increases
for model $GM3-1.31$, which has the largest disk mass at the end of the 
simulations. The reason is unknown, but might be related to a growing $m=1$ 
oscillation.

It is natural to ask if the mass of the remnant itself, excluding the disk, 
is still in the hypermassive range.
Matter which would not fall into the black hole in case of sudden collapse 
should orbit close to Keplerian speed. Hence it 
should have only little influence on the stability of the remnant at a given time.  
In Fig.~\ref{fig:evol_mdisk}, we also
show the minimum disk mass such that the remnant is supramassive. We find
that the remnant for model $GM-1.35$ is still a HMNS, for both definitions of the mass.
For the other two models, the remnant is a HMNS according to our disk measure 
$M_\mathrm{d}$, but a SMNS according to $M_\mathrm{d}^\mathrm{alt}$.

\begin{figure}[t]
\includegraphics[width=0.5\textwidth]{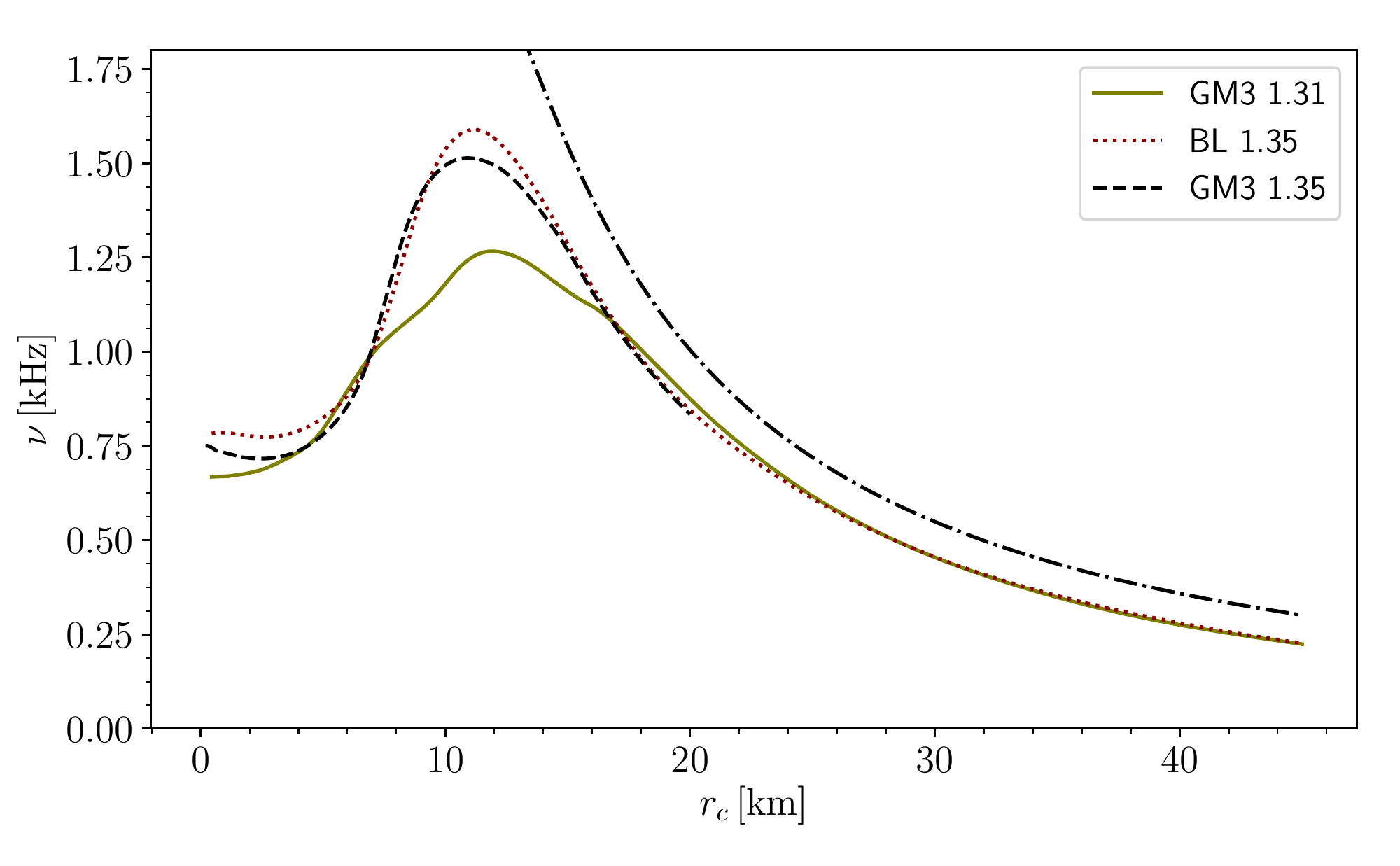}
\centering
\caption{Rotation profile of the remnant for our models $11$ ms after merger. 
$r_c$ is the circumferential radius, and $\nu$ the rotation frequency as 
observed from infinity.
The dash-dotted curve represents the 
frequency for a test particle on a prograde circular orbit.}
\label{fig:rot_prof}
\end{figure}

We now turn to discuss the structure of the HMNS remnants. Besides the mass, one of the 
properties most relevant for delaying the collapse is the rotation profile.
Figure~\ref{fig:rot_prof} shows the angular frequency as a function of 
radius $11$ ms after merger. In all cases, we observe a slowly rotating core, while the 
equatorial bulge approaches the Keplerian rotation rate. We interpret the residual 
difference in the disk as the contribution from the pressure gradient.
The shape of the rotation profile seems to be a generic feature of merger remnants,
which was observed in several studies~\cite{Kastaun:2015:064027,Endrizzi2016,
Kastaun:2016,Kastaun:2017, Hanauske:2017,Ciolfi2017} for many different different models.

We note that the profiles for model $GM3-1.31$ and $GM3-1.35$ differ significantly, despite
the similar mass.
The reason is unclear. However, the HMNSs are close to collapse and it is not surprising if
the exact evolution is very sensitive to small changes of the initial parameters.
Model $GM3-1.31$ also develops a stronger $m=1$ perturbation, which might be related.

As discussed in \cite{Kastaun:2016}, the radial mass distribution in the slowly 
rotating core can be approximated well by the core of an nonrotating NS (TOV 
solution). Since the mass of TOV solutions is bounded, this defines a critical
mass of the remnant core. In \cite{Ciolfi2017}, we proposed that the HMNS 
collapses when the TOV core equivalent reaches the maximum mass. This was 
confirmed for two models studied in \cite{Ciolfi2017}.

Before comparing the conjecture to our results, we briefly summarize 
the gauge independent measures introduced in \cite{Kastaun:2016} for this purpose.
They are based on the proper volume and baryonic mass enclosed inside 
surfaces of constant mass density. From volume and mass, we derive a radius and a 
compactness measure for each isosurface. Further, we define the bulk of a star as the region inside 
the single isosurface which is most compact according to the new measure. 
This allows us to unambiguously define the TOV core equivalent of the merger 
remnant as the TOV solution for which volume and mass of the bulk coincide with 
mass and volume of some isosurface inside the remnant. Finally, we define the 
remnant core as the interior of the isosurface with the density identical to 
the density on the TOV core equivalent's bulk surface. Inside this core, 
the radial mass distribution of equivalent TOV and remnant agree well.
For further details, see \cite{Kastaun:2016}. Bulk and core of the remnant 
for model $GM3-1.31$ are shown in Fig.~\ref{fig:gm3_disk} for comparison.

\begin{figure}[t]
\includegraphics[width=0.5\textwidth]{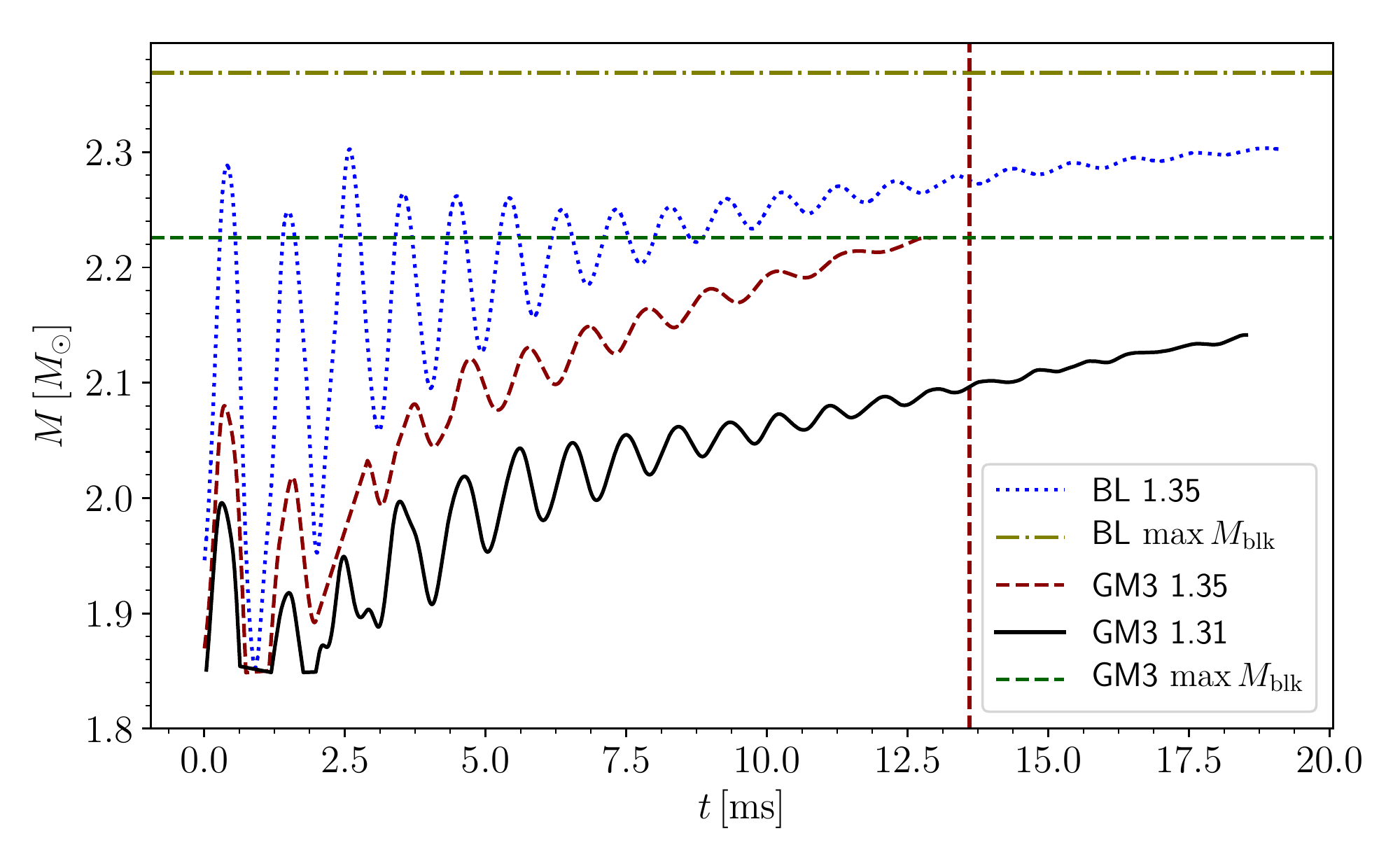}
\centering
\caption{Time evolution of the TOV solutions equivalent to the remnant cores (see text) for our simulations. 
The curves show the bulk mass of the TOV solutions, and the horizontal lines mark the maximum bulk mass for
TOV sequences of the given EOS. The vertical line marks the BH formation.}
\label{fig:eqtov_evol}
\end{figure}

The time evolution of the TOV core equivalents is shown in 
Fig.~\ref{fig:eqtov_evol}. We find that model $GM3-1.35$ indeed 
collapses once the bulk mass of the core equivalent TOV exceeds 
the maximum. 
We note that the remnant for model $GM3-1.35$ is not accreting matter 
before collapse, but, on the contrary, sheds mass into the disk.
The observed collapse could be caused by internal rearrangement 
of the HMNS structure and/or loss of the angular momentum via the 
matter migrating into the disk.

For the other models, the bulk mass is still well below 
the maximum at the end of the simulation.
Extrapolating the core evolution, we estimate that the remnants for models
$BL-1.35$ and $GM3-1.31$ might survive for another ${\approx}10\usk\milli\second$.

We stress that our discussion of remnant lifetimes is purely qualitative. 
We found that the outcome of our simulations is very sensitive to 
mass and EOS, likely because the system is close to collapse. Therefore 
we also expect the lifetime to be sensitive to numerical errors, and, 
most importantly, the physical viscosity. In \cite{Shibata:2017}, it was demonstrated 
that prescribing an effective alpha viscosity affects the evolution of the rotation profile. 
Such an effective viscosity could be caused by magnetic field 
amplification processes, although it is difficult to model the 
magnetic field on small scales numerically (see \cite{Kiuchi:2015:1509.09205}).

Last but not least, we discuss the prospects for producing a SGRB. 
In particular, we consider the scenario of a BH embedded in a disk, 
with a strong ordered magnetic field 
around the BH rotation axis.
Regardless of the disk mass definition, all our models develop a significant disk.
For model $GM3-1.35$, a BH surrounded by a disk is formed during the simulation.
For the other models, we expect collapse within tens of ms. Moreover, the disk mass is still growing
at the end of the simulations. It is therefore likely that the disk mass is not significantly reduced 
when the HMNSs finally collapse. 

We did not include magnetic fields in our simulations, and therefore do not discuss the prospects for 
strong magnetic fields in our models. For a general discussion, we refer to 
\cite{Ciolfi2017,Kawamura:2016:064012,Ruiz2016,Ciolfi2018} instead.

\begin{figure}[t]
\includegraphics[width=0.5\textwidth]{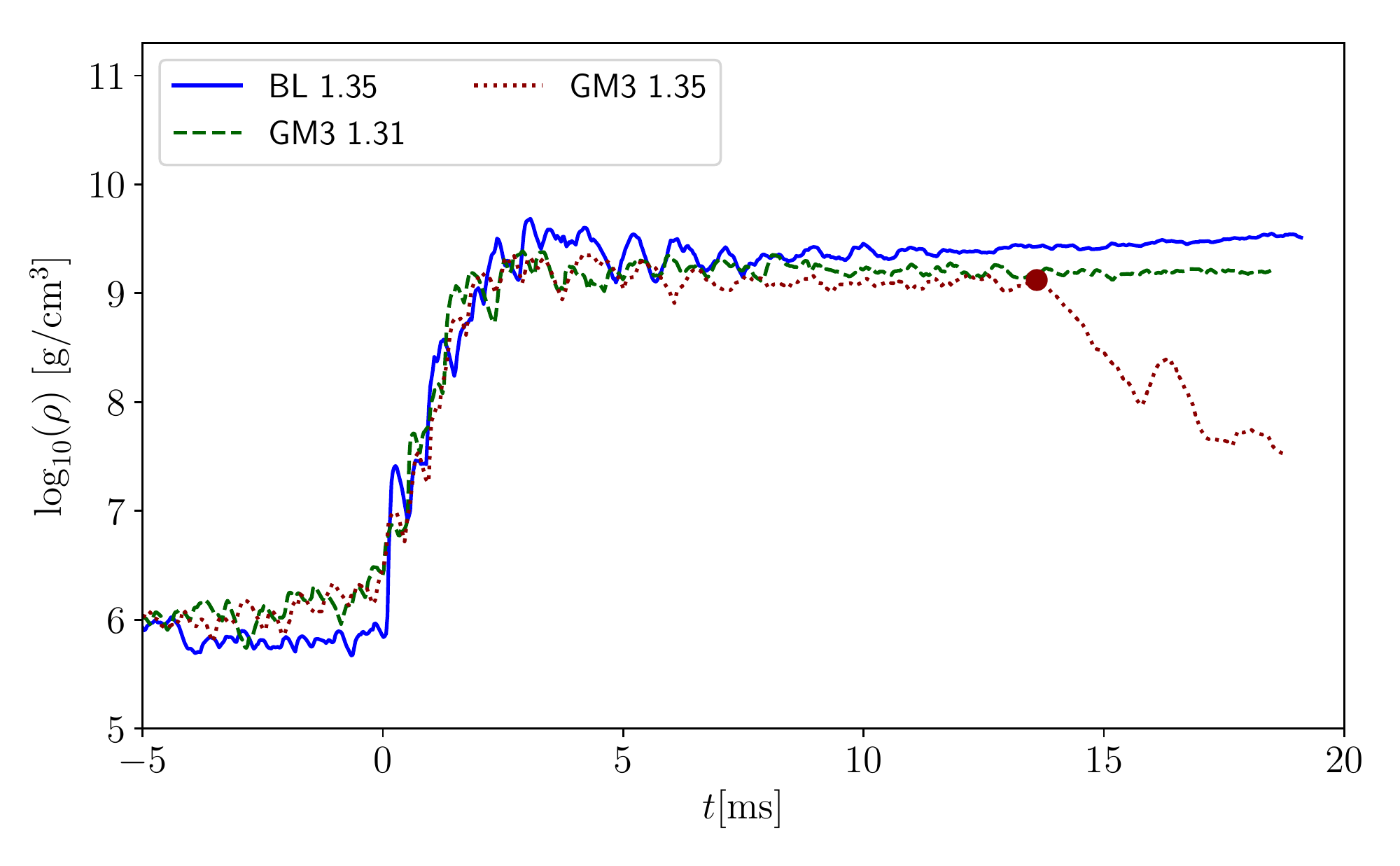}
\centering
\caption{Average rest mass density along the rotation axis above the remnant. The average is taken 
between a distance $30$ and $50$ km from the origin. The density around 
$10^6\usk\gram\per\centi\meter\cubed$ during the inspiral is due to matter ejected by numerical 
artifacts at the NS surfaces. 
For comparison, the density of the artificial atmosphere used in our simulations is 
$6\times 10^4 \usk\gram\per\centi\meter\cubed$. 
}
\label{fig:funnel}
\end{figure}

Since baryon pollution might suppress the formation of a jet, the baryon density along 
the BH axis is an important aspect with regard to SGRBs. Therefore, we computed
the rest mass density along the rotation axis, averaged between $30$--$50\usk\kilo\meter$
above the equatorial plane (see Fig.~\ref{fig:funnel}). 
We note that the density at any time after merger is more than two 
orders of magnitude above the artificial atmosphere density, which can 
be safely ignored.

The time evolution of the density is shown in Fig.~\ref{fig:rho_xz}.
During the HMNS lifetime, all models exhibit a comparable density of
${\approx}10^9\usk\gram\per\centi\meter\cubed$. For model $GM3-1.35$, the density 
decreases by almost two orders of magnitude within $5\usk\milli\second$ as 
soon as the BH is formed. 
Those findings are similar to the results obtained for different models
in a previous work \cite{Ciolfi2017}.
Our results are also relevant for studies on the possibility of 
low-mass BNS as the central engine for SGRBs (e.g., \citep{Rowlinson2013,Ciolfi:2015:36,Ciolfi2018}).
For further discussions, also compare \cite{Murguia2014,Nagakura2014,Just2016,Murguia2017}.

\section{Mass ejection and kilonova}
\label{sec:kilonova}
We now turn to discuss the ejection of matter during and after merger. 
The mass, geometrical distribution, expansion velocity, and composition of this outflow are the key ingredients to predict the associated electromagnetic emission, in particular the kilonova signal. 

Matter can be expelled dynamically during or shortly after merger by different mechanisms.
Tidal forces during merger eject cold, neutron-rich matter in the orbital plane, which 
can however be heated when faster tidal ejecta collide with slower ones.
Shock waves formed during merger can liberate matter from the remnant surface (breakout shocks).
Breakout shocks are thought to yield less neutron-rich matter, which can also be launched 
outside the orbital plane, including the rotation axis.
Oscillations of the remnant can cause waves in the surrounding debris material which can also steepen into shocks.
Since the r-process nucleosynthesis and the resulting kilonova signal depend on the thermal history, the electron fraction, and the velocity of the ejecta, it is very important to identify the different ejecta components and the corresponding ejection mechanisms (see \cite{Martin:2018} for a discussion).

To determine if a fluid element will be dynamically ejected, we adopt the geodesic 
criterion $u_t<-1$, where $u$ is the 4-velocity (see \cite{Endrizzi2016,Ciolfi2017} for 
further explanation). 
The snapshots shown in Figures~\ref{fig:rho_xy} and \ref{fig:rho_xz} also mark 
the regions with unbound matter. We observe typical spiral patterns of tidal ejecta 
in the orbital plane as well as other ejecta in all directions, likely caused by shocks.

\begin{figure}[t]
 \includegraphics[width=0.5\textwidth]{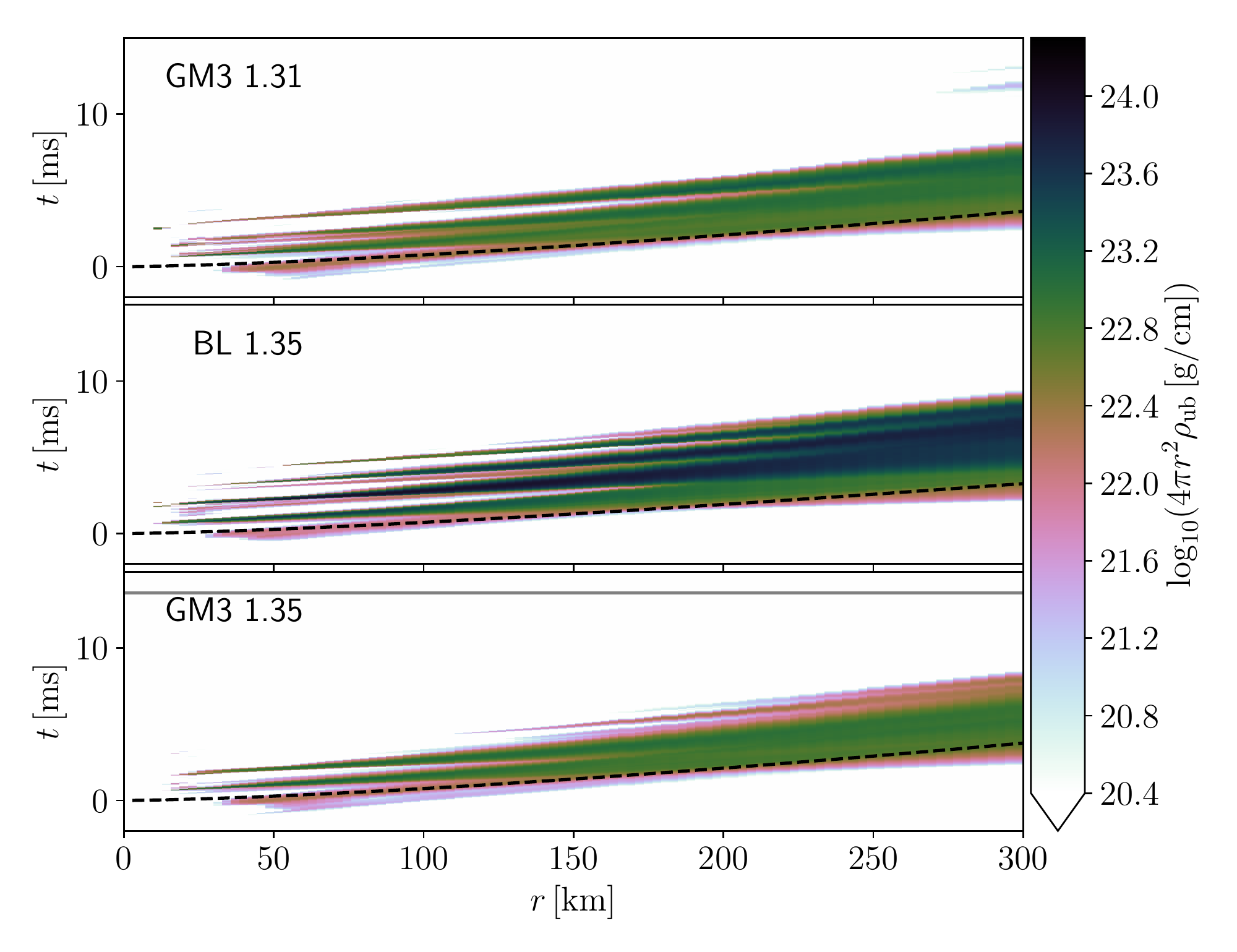}
 \centering
 \caption{Radial distribution of unbound matter as a function of time. 
 The logarithmic color scale corresponds to the unbound mass per coordinate radius. 
 The black dashed line represents the trajectory of a radially outgoing 
 test mass with velocity $v_\infty$ from Table \ref{tab:outcome}. 
 The trajectory has been estimated using the Newtonian potential of a mass 
 equal to the ADM mass at the end of the simulation.}
 \label{fig:ejecta}
\end{figure}

In Fig.~\ref{fig:ejecta} we show the time evolution of the radial distribution of unbound matter 
for the three BNS models. 
One can clearly distinguish several waves of ejected matter, which merge at larger radii.
Only the first wave is launched around the time of merger. The subsequent waves are likely 
related to the oscillations of the remnant interacting with the surrounding matter.

In order to quantify the matter ejection, we monitor the flux of unbound matter through several
spherical surfaces, at radii $50, 100, 200, 500, 700 \usk M_\odot$.
For each, we integrate the flux of unbound mass in time, and take the maximum over the different surfaces
as best estimate for the true amount of dynamically ejected matter (see 
\cite{Endrizzi2016,Ciolfi2017}). The results for all models are given in Table~\ref{tab:outcome}.

Further, we compute an average final expansion velocity from
\begin{align}
\bar{u}_t &= \frac{1}{M_e} \int  W \rho_u u_t \mathrm{d}V &
M_e &= \int W\rho_u \mathrm{d}V, 
\end{align}
where $\rho_u$ is the rest-frame (baryonic) mass density of unbound matter, 
$W$ its Lorentz factor, and $\mathrm{d}V$ the proper 3-volume element.
The integration is carried out over the region $r>150 \usk\kilo\meter$, 
outside the strongly dynamic region, and at the time where $M_e$ becomes maximal.
We obtain an average Lorentz factor $\bar{W}_\infty = - \bar{u}_t$,
which the matter will approach at infinity.
The corresponding average expansion velocity $v_\infty$ is given in 
Table~\ref{tab:outcome}. Our values fall in the range found in simulations of other BNS systems~\cite{Endrizzi2016, Ciolfi2017, Sekiguchi:2015, Bauswein:2013:131101, Hotokezaka:2013:44026, Oechslin:2007}. We estimate that thermal effects might increase the given expansion velocity by up to $20\%$ (based on comparing Bernoulli and geodesic criteria for our simulation data).

Model $BL-1.35$ result in a significantly larger amount of unbound material 
compared to the two models with GM3 EOS. 
For model $BL-1.35$, we also find larger fluctuations of the central density and 
a stronger modulation  of the GW frequency, which will be discussed in Sec.~\ref{sec:gw}.
Stronger quasi-radial oscillations of the remnant are one possible explanation 
for the higher ejecta mass from model $BL-1.35$. 
The stronger oscillations could be related to the larger compactness of the 
initial NS of this model.

\begin{figure}[t]
 \includegraphics[width=0.5\textwidth]{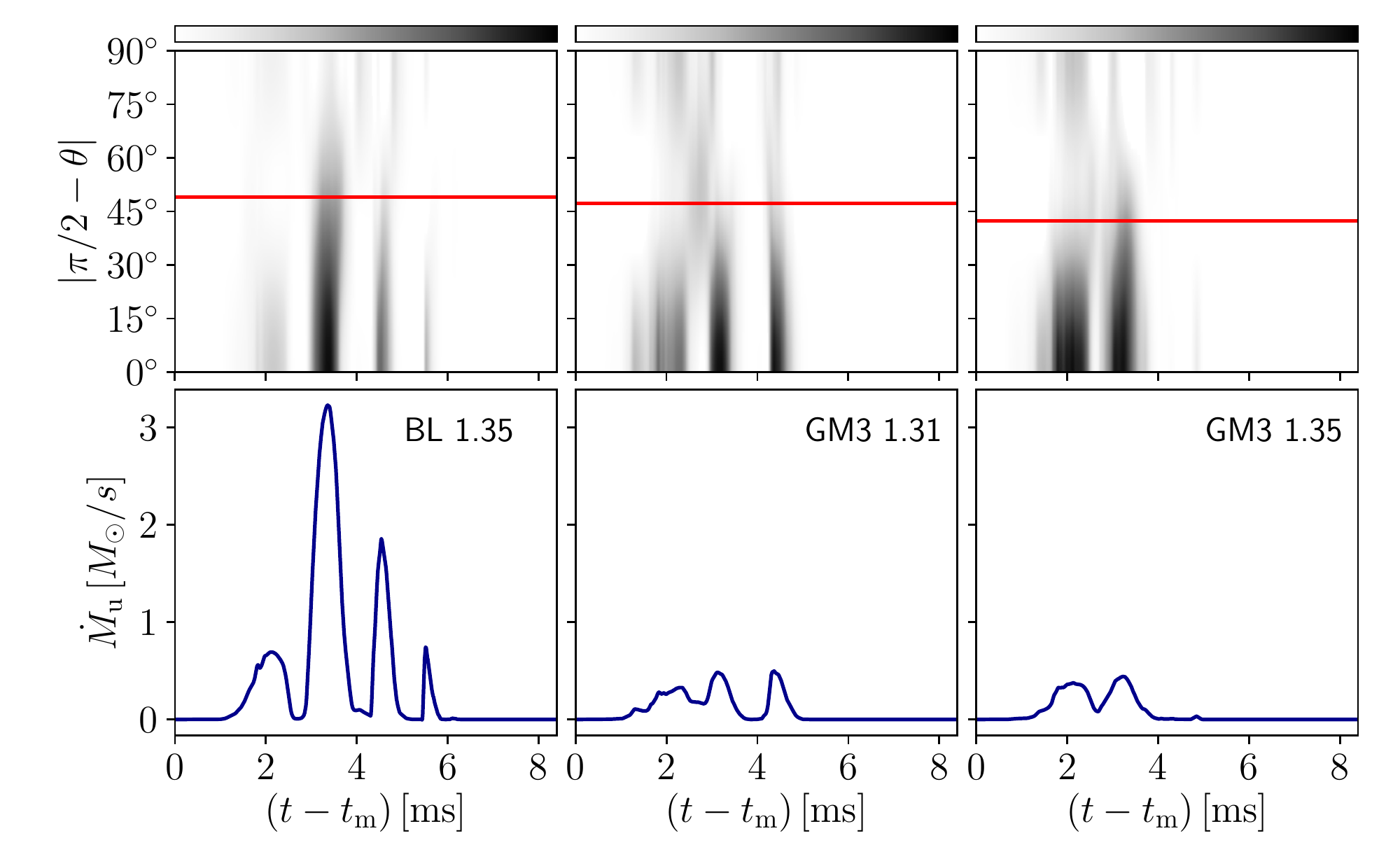}
 \centering
 \caption{Time evolution of the unbound matter flux through a spherical surface 
 at $r=148\usk\kilo\meter$.
 Top panels: polar distribution of the flux reconstructed from the multipole moments
 up to $l=4$. The orbital plane corresponds to $0^\circ$ and the orbital axis to $90^\circ$.
 The colorscale is linear, ranging from zero (white) to the maximum flux for the given 
 model (black).
 The horizontal line marks the angle below which 90\% of the matter is ejected during
 the whole evolution.
 Bottom panels: total flux through the spherical surface, showing the contribution of
 the separate ejecta waves.
 }
 \label{fig:ejecta_flux}
\end{figure}

To further investigate the main ejection mechanism, 
we compute the mass flux corresponding to 
the different waves, shown in the lower panels of Fig.~\ref{fig:ejecta_flux}.
The mass ejection for model $BL-1.35$ is indeed dominated by the waves launched after merger,
while the tidally ejected first wave contributes little.
We note that the result might change for systems with large mass ratio, 
which typically produce more tidal ejecta.

Next, we investigate the polar distribution of ejected matter.
During the simulations, we compute multipole moments (up to $l=4$) 
of the unbound matter flux on the aforementioned spherical surfaces.
From this data, we reconstruct an approximate distribution, which is
shown in the upper panels of Fig.~\ref{fig:ejecta_flux}.
We find that 90\% of the matter is ejected at angles less than $50^\circ$ 
from the orbital plane.

We can now estimate the kilonova signal that would be produced by these dynamical ejecta (ignoring 
potential later outflows).
For this, we apply the simple analytical model by \cite{Grossman2014} (using 
the same parameter value $\alpha=1.3$), and assume
that all the ejecta material is involved in the r-process nucleosynthesis (for more on kilonova models see~\cite{Li:1998,Metzger:2010,Roberts:2011,Barnes:2013,Metzger:2014}).
Besides ejected mass and expansion velocity, the analytical estimates depend strongly on the 
opacity, which depends mainly on the amount of lanthanides produced. The latter depends 
on the initial neutron fraction of the ejecta. 
Our simulations do not include neutrino radiation and employ an EOS assuming beta equilibrium and 
simplified treatment of thermal effects. Therefore, we cannot predict the ejecta composition.
Instead, we provide estimates for a plausible opacity range.
In detail, we consider lanthanide-poor ejecta with opacity 
$0.5 \usk\centi\meter\squared\per\gram$,
and lanthanide-rich ejecta with opacity
$10 \usk\centi\meter\squared\per\gram$ (see \cite{Villar2017} and references therein).

The results are given in Table~\ref{tab:kilonova}. The unknown opacity introduces
factors $4.5$, $7$, and $3.5$ to the uncertainties of peak time, peak luminosity, and 
temperature, respectively. Although a simultaneous measurement of all three would 
fix opacity, ejecta mass, and expansion velocity within the analytic model, the 
model uncertainties would likely not allow to distinguish between the BL and GM3 EOS 
for the given mass $1.35 \usk M_\odot$. However, the large differences between the 
two EOS obtained within the approximation indicate that it should be possible to distinguish 
between the two EOS considered here if the kilonova signal could be predicted more accurately. 
For the lanthanide-poor opacity, the peak time differs by a factor ${\approx}2$ between BL 
and GM3, the luminosity by a factor ${\approx}3$, and the temperature by $1.4$.

\begin{table*}
  \caption{Order of magnitude estimates for the kilonova signal caused by the dynamical ejecta, 
  computed from the analytic approximations given in \cite{Grossman2014}, 
  for two different assumptions about the opacity $\kappa$. Using ejecta mass and expansion velocities
  shown in Table~\ref{tab:outcome}, we obtain the time $t_\mathrm{peak}$ where peak luminosity 
  $L_\mathrm{peak}$ is reached, and the corresponding temperature $T_\mathrm{peak}$. We also
  provide the maximum-intensity wavelength $\lambda_\mathrm{peak}$ of the respective black-body 
  spectrum.} 
  \begin{ruledtabular}\begin{tabular}{llllll}
      Model & $\kappa \, [\centi\meter\squared\per\gram]$ & 
      $t_\mathrm{peak} \, [\mathrm{days}]$ & 
      $L_\mathrm{peak} \, [10^{40} \usk\mathrm{erg}\per\second]$ &
      $T_\mathrm{peak} \, [10^3 \usk\kelvin]$ &
      $\lambda_\mathrm{peak} \, [\nano\meter]$\\
      \hline
      \multirow{2}{*}{$BL-1.35$}  & 0.5 & 0.7 & 21   & 7.8  & 371\\
                                  & 10  & 3.0 & 3.0  & 2.3  & 1276\\  
      \multirow{2}{*}{$GM3-1.35$} & 0.5 & 0.3 & 7.8  & 11.0 & 263\\
                                  & 10  & 1.5 & 1.1  & 3.2  & 906\\
      \multirow{2}{*}{$GM3-1.31$} & 0.5 & 0.4 & 10   & 10.2 & 284\\
                                  & 10  & 1.6 & 1.5  & 3.0  & 979
  \end{tabular}\end{ruledtabular}
   \label{tab:kilonova}
\end{table*}

It is instructive to compare the potential kilonova signal of our models 
$BL-1.35$ and $GM3-1.35$ to the optical 
counterpart AT2017gfo (see \cite{GW170817-EM}) observed for the GW event GW170817.
We emphasize that there is no reason to expect agreement, since the 
chirp mass inferred for GW170817 is 1\% larger, and because the mass ratio 
is not well constrained. The evolution of luminosity and spectrum of AT2017gfo 
can be fitted by a kilonova model consisting of independent components caused by distinct
ejecta waves with different opacities, masses, and velocities \cite{Kasen2017,Villar2017,Bauswein:2013:131101}. 
The model in \cite{Kasen2017} includes a 'blue' component caused by low-lanthanide ejecta 
with mass ${\approx} 0.025 M_\odot$, expanding at ${\approx}0.3 \usk c$. 
Note the high velocity is also inferred from the fact that the observed spectrum is 
essentially featureless. Similar results are derived from the kilonova models
in \cite{Villar2017}.
Given the lower masses and lower velocities (see Table~\ref{tab:outcome}), 
we conclude that models $BL-1.35$ and $GM3-1.35$ are unlikely to produce a counterpart 
similar to the blue component of AT2017gfo, even when assuming low-lanthanide ejecta.
We also note that our ejecta are dominantly shock-driven and mostly equatorial, and it is difficult to guess their composition. Before becoming unbound, they orbit close to the remnant, while being irradiated by its neutrino emission. Further, they are heated by multiple shocks, as shown in Fig.~\ref{fig:T_v_rho} in Appendix~\ref{sec:Tfeat}.

In addition to the blue component discussed above, the kilonova signal AT2017gfo shows 
also a ``red" component, peaking in the IR band several days after merger (see, e.g., 
\cite{Kasen2017}). The latter is modeled in \cite{Kasen2017} by a very massive ejection 
of matter (${\approx}0.04 \, M_\odot$) which is expanding more slowly (${\approx} 0.1 \, c$).
The likely explanation for the high mass is that the red component of AT2017gfo was caused, 
at least partially, by winds expelled from the disk (and possibly a long-lived remnant) 
due to neutrino radiation and magnetic fields (see, e.g., 
\cite{Dessart:2009:1681,Siegel:2014:6,Siegel2017}).

Since our models possess massive disks, one should expect a contribution to the kilonova 
signal as well. The mass ejected as winds is however difficult to model, and therefore we 
can make no predictions about the corresponding lightcurves.
Instead, we discuss the implications if our models were to eject $0.04\,M_\odot$ in form 
of winds, similar to AT2017gfo. 
The first important aspect is whether the remaining mass 
could still form a BH, and how long it would take. Comparing the remaining mass to the maximum mass of 
supramassive stars (see Fig.~\ref{fig:evol_mdisk}), we find that all three models would still be 
in the hypermassive mass range, and therefore form a BH on short timescales $<1\usk\second$. 
The next question is how much of the material can originate from the disk. Comparing to 
Fig.~\ref{fig:evol_mdisk}, we find that around half of the disk mass $M_\mathrm{d}$ would need 
to be ejected for model $BL-1.35$ if the wind originates entirely from the disk. From the discussion
in Sec.~\ref{sec:dynamics}, it seems likely that a BH is formed within ${\lesssim}20\usk\milli\second$
(compare Fig.~\ref{fig:evol_mdisk}).
Therefore, we assume that winds originating directly from the surface will not contribute much.
For model $GM-1.35$, $75\%$ of the disk would need to be ejected; Since the remnant is a BH,
it cannot provide additional mass. The situation for model $GM3-1.31$ falls between the 
other two.

We caution that the above discussions are only qualitative and model-dependent. For example, 
if the merger would produce a successful (or choked) jet accompanied by a wide-angle and mildly 
relativistic cocoon (e.g., \cite{Lazzati:2018,Mooley2017}), it could alter the contribution 
of the shock-driven ejecta to the blue kilonova component, however the energy of the latter ($\sim 10^{51}\, \mathrm{erg}$) is orders of magnitude larger than the observed energies for short GRBs ($\sim 10^{49}-10^{50}
$~\cite{Berger2014}) and particularly of the jet observed for GW170817~\cite{Lazzati:2018, Margutti:2018}. In~\cite{Duffel:2018:arXiv} it is shown that jets only transfer a small amount of energy to the ejecta. Further, we ignore the ejecta 
geometry and nonuniformity in velocity and opacity (see also \cite{Villar2017,Perego2017}).


\begin{figure*}[t]
\includegraphics[width=0.95\textwidth]{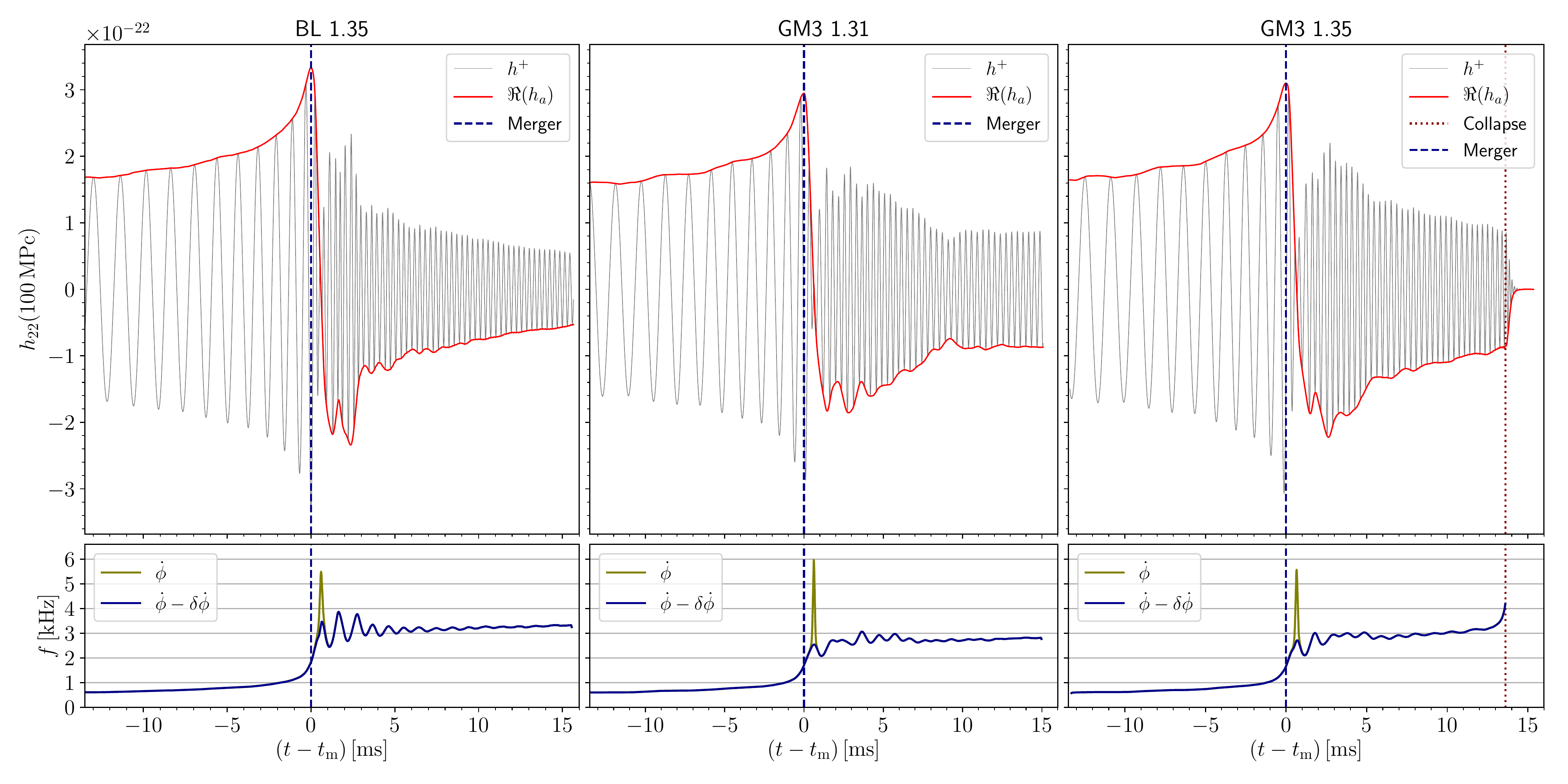}
\centering
\caption{GW strain (top panel) and instantaneous frequency (bottom panel) for our simulations. 
We only show the main contribution to the strain, given by the $l=m=2$ mode. 
The signal is scaled to a nominal source distance of $100\, \mathrm{Mpc}$. 
The gray curve is the GW strain coefficient $h^+_{22}$. 
The red curve is the generalized amplitude accounting for over-modulation (see text).
The yellow curve shows the instantaneous frequency computed from the phase of the strain, 
while the blue line shows the frequency without the contribution of the phase jump. 
The vertical dashed lines mark the merger times.
The dotted vertical lines in the right panels mark the formation of an apparent horizon.}
\label{fig:gw_strain}
\end{figure*}

\section{Gravitational Wave emission}
\label{sec:gw}
For all our simulations, we extract the GW signal at a fixed radius of $d \sim 1034\ \mathrm{km}$ using the Weyl formalism, with no extrapolation to infinity. In detail, we decompose the Weyl scalar $\Psi_4$ into spin-weighted spherical harmonics $|_{-2}Y_{lm}(\theta, \phi)|$. Integrating twice in time yields the gravitational strain coefficients 
$h_{lm} = h_{lm}^+ - i h_{lm}^\times$. The time integrations are performed using the method described 
in~\cite{Kastaun:2016}. It effectively suppresses minor low-frequency offsets of numerical origin, 
which would otherwise get amplified in the integration.

GW strain and phase velocity for the three models are shown in Figure~\ref{fig:gw_strain}. 
Only for model $GM3-1.35$, a BH was formed during the simulation, and one can see the 
corresponding decay of the amplitude around $14\,\mathrm{ms}$ after merger. 
From mass and spin of the BH produced for model $GM3-1.35$, we obtain a ringdown 
frequency $f_\mathrm{BH} = 6.52 \usk\kilo\hertz$.

Comparing models $BL-1.35$ and $GM3-1.35$, which have identical NS masses, we find a moderate influence 
of the EOS. The most prominent difference is the length of the post-merger GW signal, 
given by the remnant lifetime. For model $BL-1.35$, the remnant survives at least 
$5\,\mathrm{ms}$ longer (the corresponding GW signal is not fully visible in the 
figure since it needs time to reach the extraction radius). As discussed in 
Section~\ref{sec:dynamics}, the collapse is probably delayed even longer.

The peak amplitude at merger is slightly larger for model $BL-1.35$. Subsequently, the 
amplitude decays faster for this model compared to model $GM3-1.35$. Another difference 
is that the frequency modulation, visible during the first few ms 
after merger, is stronger for model $BL-1.35$. 
Finally, the post-merger frequency is slightly higher for model $BL-1.35$, except close to 
the BH formation for model $GM3-1.35$. 

Both the larger amplitude and stronger frequency modulation for model $BL-1.35$ 
might be related to the fact that the initial NSs are more compact (see 
Table~\ref{tab:init_param}), such that the stars come into contact later and at higher velocities.

Comparing the two models with GM3 EOS, we find that the small mass difference has a significant
impact on the length of the signal: the lighter model $GM3-1.31$ survives at least $5\usk\milli\second$ 
longer, probably more (see Section~\ref{sec:dynamics}). 
The merger- and post-merger amplitude is slightly larger for model $GM3-1.35$, and also the 
post-merger frequency is slightly higher for the latter. Both models show only weak modulation of the
instantaneous frequency after merger. Model $GM3-1.35$ exhibits a faster drift towards higher 
frequencies, which is probably related to the shorter lifetime.

We also employ a scheme to detect phase jumps caused by over-modulation. The latter describes a signal
$A(t)e^{i\phi(t)}$, with a slowly changing amplitude $A\in \mathbb{R}$ that is allowed to 
cross zero. For the case of GW signals, this could happen if the dominant multipole moment of the remnant in a 
co-rotating frame changes sign. Further details of the method can be found in~\cite{Kastaun:2016}.

As in \cite{Kastaun:2016, Ciolfi2017}, we find that the strain minima occurring in our simulations 
directly after merger are likely caused by over-modulation. The generalized amplitude (obtained by 
allowing negative values, but no sudden phase jump, see \cite{Kastaun:2016}) is shown in the upper 
panels of Figure~\ref{fig:gw_strain}. The lower panels shows that the sharp peaks in the phase 
velocity after merger (yellow curve) vanish when removing the contribution from 
over-modulation (blue curve).

\begin{figure}[t]
\includegraphics[width=0.5\textwidth]{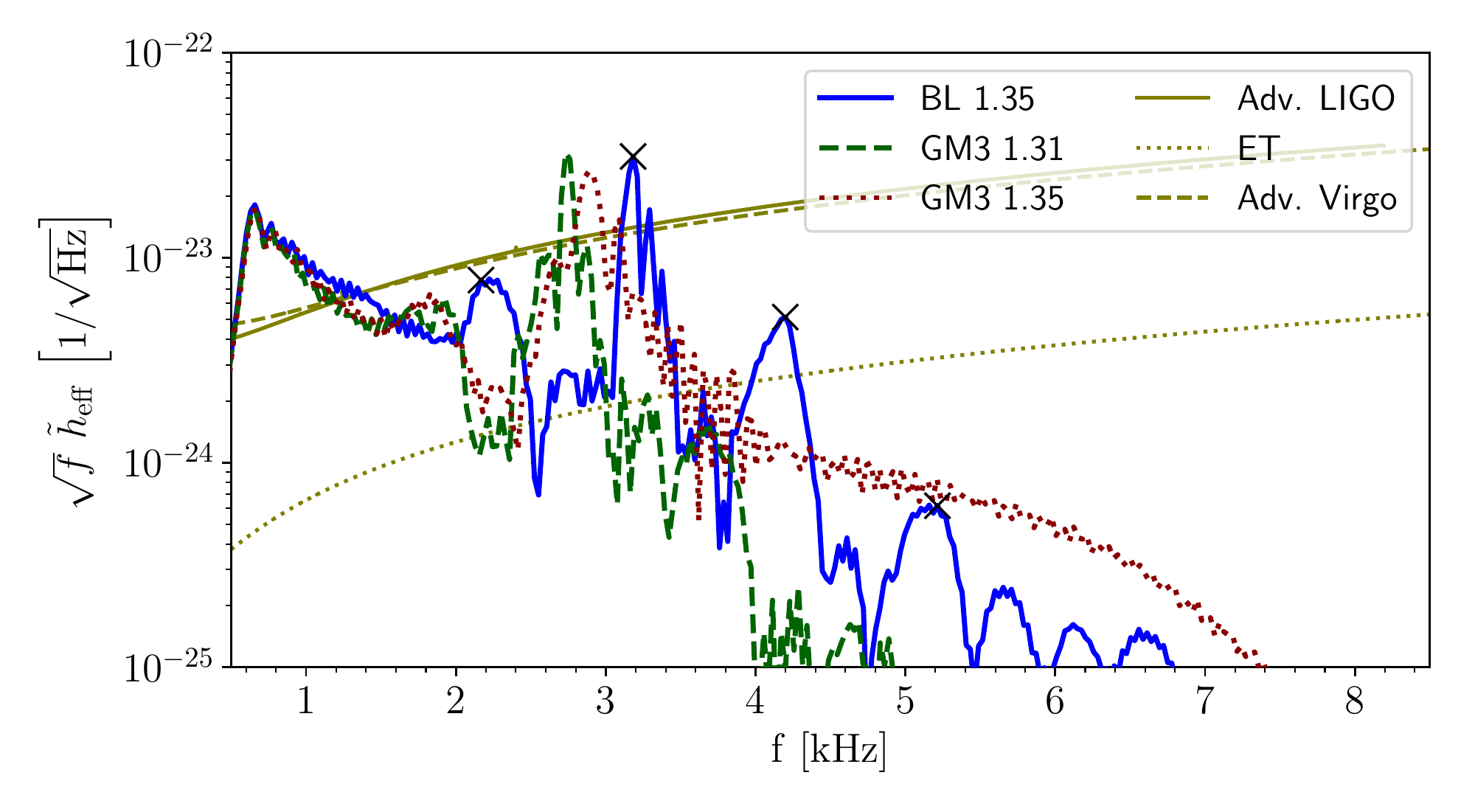}
\centering
\caption{Power spectrum of the GW strain for our simulations plotted together with the sensitivity curves for advanced LIGO and Virgo and the planned Einstein Telescope. The source is assumed at a distance $100\,\mathrm{Mpc}$ from the detector, 
oriented with orbital angular momentum along the line of sight. The symbols mark the frequencies of the main peak
and combination frequencies $f_\mathrm{pm} + n f_\rho$, where $n=-1,1,2$ and $f_\rho = 1.0\usk\kilo\hertz$ is 
the quasi-radial oscillation frequency.}
\label{fig:gw_spec}
\end{figure}

The GW power spectra for the three models are shown in \Fref{fig:gw_spec}.
For each, the dominant post-merger peak is clearly visible. 
Their frequencies are given in Table~\ref{tab:outcome} and correspond to 
the instantaneous frequencies after merger discussed above.

To assess the detectability of the post-merger signal with the
current LIGO detectors, we compare to \cite{GW170817-postmerger}. 
The latter study compared several existing numerical relativity waveforms (which included
the post-merger phase) to the instrument sensitivity during the detection of
event GW170817. Assuming the same sky location and distance inferred for GW170817,
the post-merger signal amplitudes were around one order of magnitude too low
for detection (see \cite{GW170817-postmerger} for details).

The relevant quantities defined in \cite{GW170817-postmerger} 
are a mean amplitude $h_{rss}$ and mean frequency $\bar{f}$ in the frequency
range ${~}1$ -- $4\usk\kilo\hertz$. 
Note this range also includes part of the merger signal.
For our model $BL-1.35$,
we find $h_{rss} = 0.26\times 10^{-22}/\sqrt{\hertz}$ and 
$\bar{f} = 2.36 \usk\kilo\hertz$. Those are typical values compared to the cases 
studied in \cite{GW170817-postmerger} (note however that we slightly underestimate the post-merger peak amplitude, 
since the $BL-1.35$ simulation ends when the GW amplitude is still large). 

For comparison, \Fref{fig:gw_spec} also shows the sensitivity curves of future detectors.
Those indicate that the post-merger signals for our models  at $100$ Mpc might be marginally 
detectable with advanced LIGO at design sensitivity, and should be detectable with the 
Einstein Telescope.

Besides the dominant post-merger peak, only model $BL-1.35$ exhibits significant secondary peaks.
Despite the low amplitude, those peaks might be detectable with the future Einstein Telescope.

In principle, secondary peaks can have different interpretations. First, they could directly 
correspond to different oscillation modes excited in the remnant. Second, quasi-radial 
oscillation can modulate the amplitude of the main $m=2$ mode (note that in a hypermassive NS, 
the radial oscillation frequency is typically lower than the $m=2$ frequency). 
The amplitude modulation then causes combination frequencies in the power spectrum, which 
differ from the frequency of the dominant $m=2$ mode by integer multiples of the quasi-radial 
oscillation frequency~\cite{Stergioulas:2011:427}. Finally, changes in the remnant structure can affect the frequency 
of the dominant $m=2$ mode as well, causing broadening or splitting of the main 
peak (see Fig. 20 in \cite{Kastaun:2017} for an example).

For the case of model $BL-1.35$, the secondary peaks
are caused by amplitude modulation. This can be seen in \Fref{fig:gw_spec}, marking the main frequency
$f_\mathrm{pm}$ as well as combination frequencies with the radial oscillation frequency 
$f_\rho = 1.0\usk\kilo\hertz$ (obtained from the power spectrum of the maximum density evolution).
In \cite{Kastaun:2017}, we noted that low frequency post-merger peaks can change strongly when removing the phase-jump
during merger, due to cancellation effects with the late inspiral in the same frequency range.
This is also the case here; however, we verified that the low-frequency peak is still present in 
the part of the signal starting at the phase-jump.

\section{Conclusions}
\label{sec:conclusion}
We presented simulations of binary neutron star mergers with  
the recently introduced BL EOS, which is derived from chiral 
effective field theory. For comparison, we carried out another 
simulation employing the GM3 EOS. 
Both evolve a system with equal mass NSs of $1.35 M_{\odot}$ each. 
To assess the influence of the mass, we study a third model with GM3 EOS 
and slightly lower NS masses of $1.31 M_{\odot}$.

Neither EOS includes thermal effects, while they both assume beta-equilibrium.
We approximated the temperature dependence using a simple gamma law, 
and ignore composition effects (a finite temperature version of the BL EOS 
will be the subject of future work). We also caution that our simulations 
do not include magnetic fields, and effective viscosity caused by magnetic fields
or turbulence. This could influence the lifetime of the HMNS and the disk mass.

All models result in a hypermassive remnant, although only the heavier GM3 model 
formed a BH within the duration of our simulations, around $14 \usk\milli\second$
after merger. Not surprisingly, the mass had a strong impact on the lifetime. 
The lighter models survive at least $5\usk\milli\second$ longer.
The collapsing model supports a recent conjecture  that the remnant 
mass distribution resembles a nonrotating NS in the core, and that HMNSs 
collapse once the corresponding TOV star reaches the maximum mass \cite{Ciolfi2017}. 

Both mass and EOS had a strong impact on the ejected mass.
The BL EOS model ejects ${\approx}6$ times more mass than the GM3 model with 
the same gravitational mass and ${\approx}4$ times more than the lighter 
GM3 model. The reason might be the larger compactness of the initial NS 
with BL EOS, resulting in a later merger at larger velocities. In any 
case, we find that the BL model shows stronger oscillations after merger,
and that most of the matter is ejected in several waves after the merger.

For all models, 90\% of the mass is ejected within $50\degree$ around the
orbital plane. Due to our simplified treatment of the matter, we cannot 
estimate the composition of the ejecta, which have a large influence on 
the corresponding kilonova lightcurves. Using a standard range of opacities 
and order of magnitude analytical estimates based on ejecta mass and velocity, 
we find however that our models likely produce very different lightcurves. 
Accurate kilonova models would allow to distinguish the two EOSs from 
observations of the EM counterparts, provided that the masses are known.
Comparing to kilonova models \cite{Kasen2017} describing the recent event 
GW170817, we find that the dynamical ejecta from our models are likely too 
slow and not massive enough to produce a similar contribution to the luminosity 
and spectral evolution.

The difference between the two EOSs is also evident in the post-merger
GW signal. The main post-merger frequency for the $1.35\usk M_\odot$ systems  
is around 10\% higher for the BL EOS than for the GM3 EOS. However, the mass
also has a strong impact. The lighter GM3 model (with ${\approx}3\%$ lower 
total baryon mass) results in 5\% lower frequency. Detecting the post-merger 
signal would therefore clearly distinguish between the two EOSs considered 
here, provided the inspiral signal allows an accurate measurement of the total 
mass. 
At a fiducial distance of 100 Mpc, the postmerger peaks will be 
barely detectable with advanced LIGO at design sensitivity, but well resolved
by the planned Einstein Telescope.

\begin{figure*}[ht]
 \includegraphics[width=\linewidth]{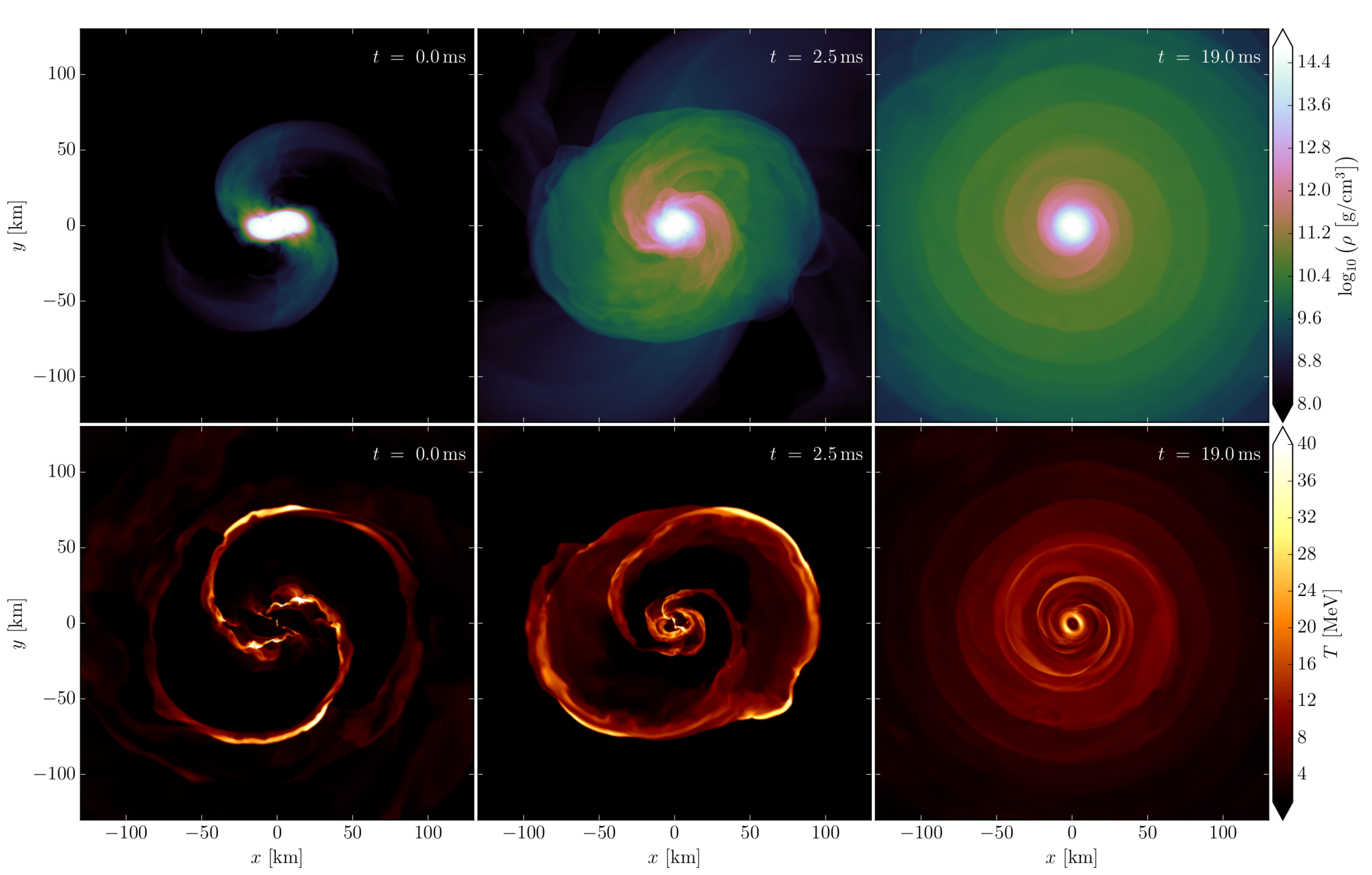}
 \caption{Snapshots of the rest-mass density (top) and temperature (bottom) in the equatorial 
 plane, for model BL-$1.35$ at merger time (left), $2.5$ ms after merger (center), and $19$ ms after merger (right).}
 \label{fig:T_v_rho}
\end{figure*}

Due to the residual eccentricity and low number of orbits, we did not measure 
the tidal effects in our simulations, but we did compute the tidal deformabilities 
of our BL and GM3 NS models. Based on those, we expect that even the current 
LIGO detectors could likely distinguish between BL and GM3 EOS from the 
inspiral GW signal alone (compare~\cite{GW170817-detection}). 

Last but not least, we studied the masses of the debris disks surrounding 
the final BH. We compared two different definitions of disk mass. One is
based on a cutoff density before BH formation, using the same value as 
in \cite{Radice2018}, and the other on a cutoff radius. We calibrated the 
cutoff radius to the mass remaining outside the BH for the collapsing model, 
and found that the density-based definition overestimates the final disk 
mass by around 60\% for this model. Further, we found that the disk is
not accreting on the remnant during the lifetime of the HMNSs, but instead 
increases due to mass shed by the remnant.
Those ambiguities of disk mass could become relevant when estimating the potential 
contribution of disk winds to the kilonova signal, although the maximum 
fraction of ejected disk mass adds a larger uncertainty.


\section*{Acknowledgments}

We acknowledge PRACE for awarding us access to Marconi at CINECA, Italy (grant 2016153613). Numerical calculations have been made also possible through a CINECA-INFN agreement, providing access to resources on MARCONI at CINECA. We also acknowledge the CINECA award under the ISCRA initiative, for the availability of high performance computing resources and support (grant IscrC\_BNSTHM1).


\appendix

\begin{figure}[ht]
 \includegraphics[width=\linewidth]{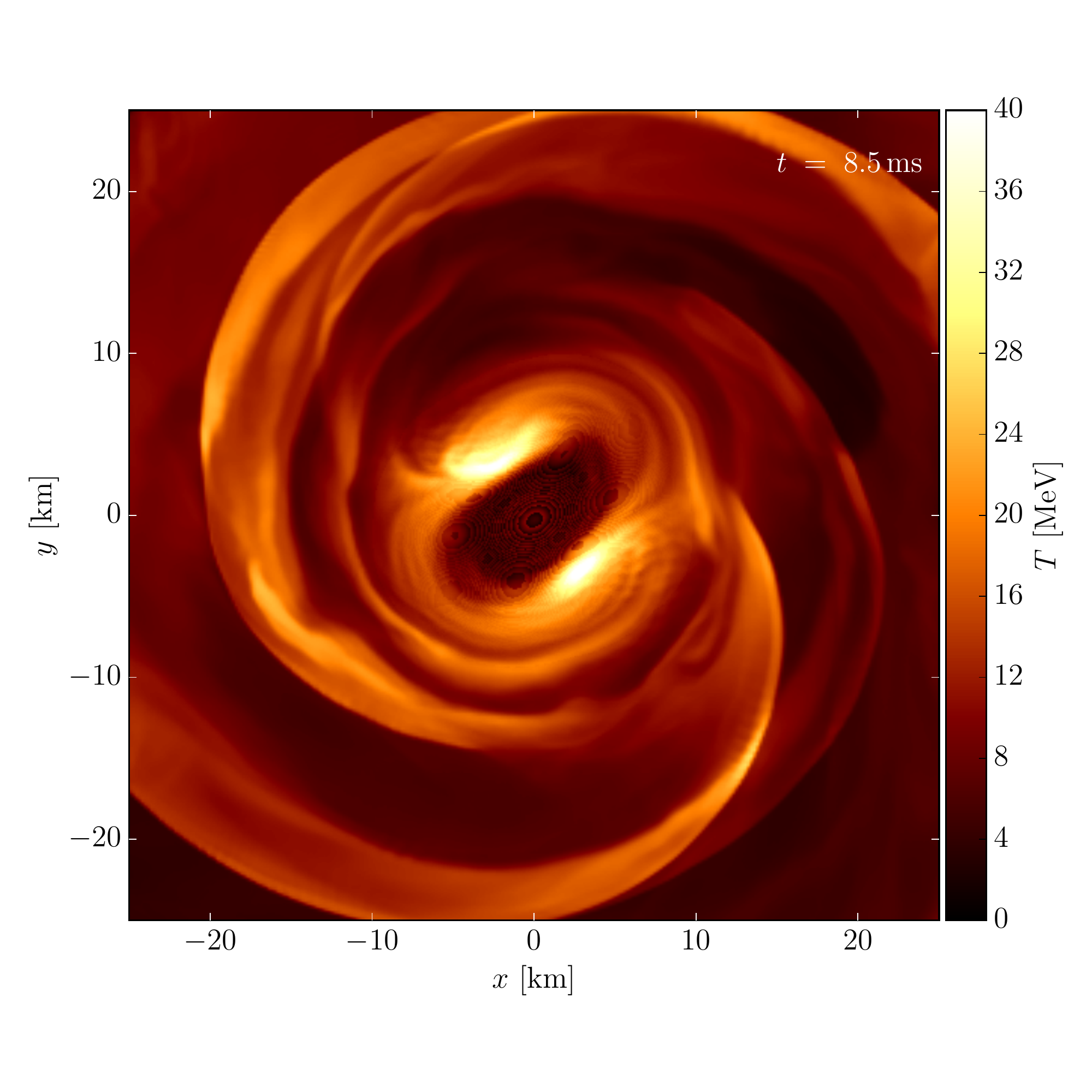}
 \caption{Temperature inside the HMNS on the equatorial plane, $8.5$ ms after merger, 
 for model BL-$1.35$.}
 \label{fig:T_snap}
\end{figure}

\section{Temperature Evolution}
\label{sec:Tfeat}
For the simulations in this work, we added thermal contributions to the cold BL and GM3 EOS
by prescribing a simple gamma law of the form
\begin{equation}
  P(\rho, \epsilon) = P_{cold}(\rho) + (\Gamma_{th} - 1) (\epsilon - \epsilon_{cold})\rho
  \label{eq:Tgammalaw}
\end{equation}
where $P_{cold}$ and $\epsilon_{cold}$ are given by the cold EOS tables (as functions of $\rho$), 
$\epsilon$ is the specific internal energy evolved by our code, and $\Gamma_{th}=1.8$.  

In order to study the thermal evolution in our simulations, we extract the specific thermal energy 
$\epsilon_{th} = \epsilon - \epsilon_{cold}(\rho)$. 
In analogy to the ideal mono-atomic gas, we define a temperature 
\begin{equation}
  T =  \frac{2}{3} \frac{m_b}{k_b} \epsilon_{th}
  \label{eq:T_extr}
\end{equation}
where $k_b$ is the Boltzmann constant and $m_b$ the rest-mass of a nucleon. 

In Fig.~\ref{fig:T_v_rho}, we show the evolution of rest-mass density $\rho$ and temperature $T$ defined above on the equatorial plane during and after merger, for model $BL-1.35$. One can clearly see spiral shocks which heat the disk and the dynamical ejecta. The heating of the disk continues during the lifetime of the HMNS, a result we also observed in \cite{Kastaun:2016} for a different model. In previous works ~\citep{Kastaun:2016, Kastaun:2017}, employing a fully tabulated EOS including finite-temperature and composition effects, we observed a characteristic thermal pattern with two hot spots in the post-merger remnant. The models studied in this work exhibit a very similar structure. This is shown in Fig.~\ref{fig:T_snap}, which can be compared to Fig.~12 in~\citep{Kastaun:2016}.

%


\end{document}